\documentclass[10pt,journal]{IEEEtran}

\AtBeginDocument{%
  \providecommand\BibTeX{{%
    \normalfont B\kern-0.5em{\scshape i\kern-0.25em b}\kern-0.8em\TeX}}}

\usepackage{inputenc}
\usepackage{subcaption}
\usepackage{graphicx}
\usepackage[english]{babel}
\usepackage{lipsum}
\usepackage{textcomp}
\usepackage{multicol}
\usepackage{hyperref}
\usepackage{array}
\usepackage{tabularx}
\usepackage{amsmath,amssymb,amsfonts}
\usepackage{placeins}
\usepackage{ctable}
\usepackage{caption}
\usepackage{multirow}
\usepackage{float}
\usepackage{makecell}
\usepackage[ruled,vlined, linesnumbered]{algorithm2e}
\usepackage{siunitx}
\graphicspath{ {images/} }

\usepackage{soul}
\usepackage{ragged2e}

\usepackage{environ}
\usepackage[font=footnotesize]{caption}

\usepackage{blindtext}
\usepackage{lipsum}
\usepackage{textcomp}
\usepackage{xcolor}
\usepackage{multicol}
\usepackage{hyperref}

\usepackage{array}
\usepackage{tabularx}
\usepackage{threeparttable}

\numberwithin{equation}{section}
\newcommand{\etal}{\textit{et al.}}

\newcolumntype{L}{>{\raggedright\arraybackslash}X}

\begin{document}




\title{Efficient and Fault-Tolerant Memristive Neural Networks with {\it In-Situ} Training
}



\author{Santlal Prajapati,
        Manobendra Nath Mondal,
        and Susmita Sur-Kolay
\IEEEcompsocitemizethanks{\IEEEcompsocthanksitem S. Prajapati, School of Engineering,Sister Nivedita University, DG 1/2 New Town, Action Area 1, Kolkata - 700156, E-mail: santlal.p@snuniv.ac.in

\IEEEcompsocthanksitem S. Sur-Kolay, ACM Unit, Indian Statistical Institute, Kolkata-700108, India, ssk@isical.ac.in

\IEEEcompsocthanksitem M. N. Mondal is now with School of Computer Science, UPES, Dehradun, India, E-mail: manobmondal@gmail.com}
\thanks{All three authors were in Indian Statistical Institute, Kolkata-700108, India, when this work was  done. An earlier version of this paper appeared in the Proc. of SOCC 2022~\cite{santlal2022}}}





\maketitle

\begin{abstract}
Neuromorphic architectures, which incorporate parallel and in-memory processing, are crucial for accelerating artificial neural network (ANN) computations. This work presents a novel memristor-based multi-layer neural network (memristive MLNN) architecture and an efficient in-situ training algorithm. The proposed design performs matrix-vector multiplications, outer products, and weight updates in constant time $\mathcal{O}(1)$, leveraging the inherent parallelism of memristive crossbars. Each synapse is realized using a single memristor, eliminating the need for transistors, and offering enhanced area and energy efficiency. The architecture is evaluated through LTspice simulations on the IRIS, NASA Asteroid, and Breast Cancer Wisconsin datasets, achieving classification accuracies of 98.22\%, 90.43\%, and 98.59\%, respectively. Robustness is assessed by introducing stuck-at-conducting-state faults in randomly selected memristors. The effects of nonlinearity in memristor conductance and a 10\% device variation are also analyzed. The simulation results establish that the network's performance is not affected significantly by faulty memristors, non-linearity, and device variation.



\end{abstract}











\begin{IEEEkeywords}
Neuromorphic computing,  memristor, memristive synapse, memristive neural network, {\it in-situ} training.
\end{IEEEkeywords}




\section{Introduction}
A memristive neuromorphic computing system excels in real-time data processing due to its capabilities of parallel and in-memory computation, which overcome the memory-wall bottleneck of von Neumann architectures~\cite{wulf_1995_ACM_hitting,mead1990neuromorphic}. Memristors, as non-linear passive circuit elements, are promising synaptic devices in neuromorphic circuits because of their tunable nonvolatile conductance states, similar to biological synapses. The conductance of a memristor in a memristive synapse represents synaptic weight~\cite{jo2010nanoscale}. Various memristor-based artificial synapses have been proposed, including memristor bridges~\cite{adhikari2014ITCS,tarkov2016OMNN}, two transistors and one memristor (2T1M)~\cite{shahar15}, one transistor and one memristor (1T1M)~\cite{ETCI_Shi_2021}, two memristors (2M)~\cite{li_2018_nature_communication, CS1RP_Okrestinskaya_2019,santlal2022}, and one memristor (1M)~\cite{sheridan2017nature_nanotechnology,CS1RP_Yzhang_2018}.  

 Memristive multi-layer neural networks (memristive MLNNs) have been constructed by using these synapses on memristive crossbars (MCBs)~\cite{Linn_2014_ISCAS}. These MCBs act as memory by storing synaptic weight matrices, and also as in-memory processors by performing matrix-vector multiplication~\cite{Li_IMW_2018}, vector-vector outer product~\cite{shahar15}, and weight matrix updates~\cite{shahar15} on the MCBs.

Memristive MLNNs~\cite{tian_2013_chinese_physics,santlal2022, yao_2017_Nature_communication, prezioso_2015_Nature, lin_2016_Scientific_reports, CS1RP_Yzhang_2018,shahar15, CS1RP_Okrestinskaya_2019, ETCI_Shi_2021,yakopcic_2015_IJCNN, li_2018_nature_communication} are trained either {\it ex-situ} or {\it in-situ}. In {\it ex-situ} training, mirror neural networks are trained on external circuits or software, and the final weights are set in the MCBs of memristive MLNNs~\cite{tian_2013_chinese_physics, alibart_2013_nature_communication, yakopcic_2015_IJCNN}. The {\it in-situ} training is performed directly on the neuromorphic hardware, which is more energy-efficient and faster due to reduced communication with external hardware~\cite{alibart_2013_nature_communication}. Moreover, it overcomes hardware imperfections~\cite{li_2018_nature_communication} and is thus a suitable alternative to address memristor imperfections and training latency.

For memristive MLNNs, the key factors include the size of artificial synapses and the time required to modify weights during in-situ training. The power consumption and physical area of transistors are much greater than those of memristors~\cite{Elshamy2015vlsi} which encourages avoiding using transistors in synapses. In~\cite{Wen_2024_ITIE}, a perceptron neural network is proposed where each synapse consists of four memristors. The studies in ~\cite{shahar15, yan2020NN} use two transistors per synapse (2T1M type) but update the conductances of all the memristors in an MCB in constant time. The works in \cite{ETCI_Shi_2021, Renhai_TCAS_2023} have reduced the MCB size of MLNNs to 1T1M synapses, achieving a constant-update time of MCB during in-situ training. For space and energy efficiency, the 2M type synapses without transistors have been proposed in~\cite{li_2018_nature_communication, CS1RP_Okrestinskaya_2019}. But during training, the time taken to update the memristors in the crossbar is longer, namely one crossbar line at a time in~\cite{li_2018_nature_communication} and one memristor at a time in~\cite{CS1RP_Okrestinskaya_2019}. The memristive MLNN~\cite{santlal2022} with 2M synapse achieved constant time update of the weight matrix during in-situ training. However, to reduce the size of the MCB (1M type), Zhang \etal~\cite{CS1RP_Yzhang_2018} proposed an artificial synapse with only one memristor, but their training method updates the conductances of the memristors of an MCB one by one, making the training time-inefficient.

Our proposed memristive MLNN architecture with {\it in-situ} training method focuses on reducing the memristive crossbar area (1M type) as well as improving the crossbar update time to $\mathcal{O}(1)$. Additionally, this work scrutinizes the fault tolerance, the effects of device variation, and the non-linearity present in the conductance of a memristor on the performance of memristive MLNNs.

\begin{table*}[h]
    \centering
    \caption{Qualitative comparison of our proposed work with earlier in-situ training methods.}
    \label{tab:comparison}
    \resizebox{\linewidth}{!}{%
  \begin{tabular}{|c|c|c|c|c|c|c|c|}
  \hline
  \multicolumn{8}{|c|}{BAM: bidirectional associative memory, MLNN: multi-layered neural network, BP: back propagation} \\
\hline
                     Work & Synapse type & \begin{tabular}[c]{@{}l@{}}Type of network\\ and its input\end{tabular}        & Training    & Update on weight                                                           & \begin{tabular}[c]{@{}l@{}}Memristor's weight\\ update voltage\end{tabular}  & \begin{tabular}[c]{@{}l@{}} Non-linearity \\analysis  \end{tabular} &  \begin{tabular}[c]{@{}l@{}}   Fault \\ analysis  \end{tabular}                       \\ \hline
Zhang \etal~\cite{CS1RP_Yzhang_2018}        &      1M        & \begin{tabular}[c]{@{}l@{}}With digital input, MLNN\end{tabular}           & Adaptive BP & \begin{tabular}[c]{@{}l@{}}One memristor\\  at a time\end{tabular}      & \begin{tabular}[c]{@{}l@{}}Constant voltage amplitude,\\ weight adjustment encode\\  by duration\end{tabular}  & No  &  No                  
\\ \hline
Li \etal~\cite{li_2018_nature_communication}           &    2M          & \begin{tabular}[c]{@{}l@{}}Analog input MLNN\end{tabular}            & General BP  & \begin{tabular}[c]{@{}l@{}}One memristor \\ line at a time\end{tabular} & \begin{tabular}[c]{@{}l@{}}Fixed duration time,\\ weight adjustment,\\ encode by  voltage amplitude\end{tabular} & No  & Yes                   

\\ \hline
Krestinskaya \etal~\cite{CS1RP_Okrestinskaya_2019} &      2M        & \begin{tabular}[c]{@{}l@{}}ANN with digital\\ or analog input\end{tabular} & General BP  & \begin{tabular}[c]{@{}l@{}}One memristor\\ at a time\end{tabular}       & \begin{tabular}[c]{@{}l@{}}~~~~~~~~~~~-do-\end{tabular} & No  &  No                  

\\ \hline
Shi \etal~\cite{ETCI_Shi_2021}       &       1T1M & 
\begin{tabular}[c]{@{}l@{}}BAM with\\digital and analog input\end{tabular} & 
\begin{tabular}[c]{@{}l@{}}Network specific \\ algorithm~\cite{chartier_NN_2006} \end{tabular} &
\begin{tabular}[c]{@{}l@{}}All memristors\\ at a time\end{tabular} & \begin{tabular}[c]{@{}l@{}}Fixed voltage amplitude\\ with feedback duration\\ time control\end{tabular}
&  No &   No                 
\\
\hline
Feng \etal~\cite{Renhai_TCAS_2023}       &       1T1M & 
\begin{tabular}[c]{@{}l@{}}MLNN with analog input\end{tabular} & 
\begin{tabular}[c]{@{}l@{}}General BP \end{tabular} &
\begin{tabular}[c]{@{}l@{}}All memristors\\ at a time\end{tabular} & \begin{tabular}[c]{@{}l@{}}~~~~~~~~~~~-do-\end{tabular}
& No  &   No                 
\\ \hline

Soudry \etal~\cite{shahar15}    & 2T1M         & \begin{tabular}[c]{@{}l@{}}Analog input MLNN\end{tabular}            & General BP  & \begin{tabular}[c]{@{}l@{}}All memristors\\ at a time\end{tabular}       & \begin{tabular}[c]{@{}l@{}}~~~~~~~~~~~~-do-\end{tabular}  
& No  &    No                
\\ \hline

Yan \etal~\cite{yan2020NN}    & 2T1M         & \begin{tabular}[c]{@{}l@{}}Analog input MLNN\end{tabular}            & \begin{tabular}[c]{@{}l@{}}Momentum and \\ adaptive learning \end{tabular} & \begin{tabular}[c]{@{}l@{}}All memristors\\ at a time\end{tabular}       & \begin{tabular}[c]{@{}l@{}} ~~~~~~~~~~~~-do- \end{tabular}    &  No & No                   
\\ \hline

Prajapati \etal~\cite{santlal2022}         & 2M           & Analog input MLNN                                                                      & General BP  & \begin{tabular}[c]{@{}l@{}}All memristors\\ at a time\end{tabular}       & \begin{tabular}[c]{@{}l@{}}Variable voltage amplitude\\ with feedback duration\\ time control\end{tabular} 
 & No  & No                   
\\ \hline

{\bf Proposed work}         & {\bf 1M}           &{\bf Analog input MLNN}                                                                      & {\bf General BP}  & \begin{tabular}[c]{@{}l@{}}{\bf All memristors}\\ {\bf at a time}\end{tabular}       & \begin{tabular}[c]{@{}l@{}}~~~~~~~~~~~~{\bf -do-}
\end{tabular}    &  {\bf Yes} & {\bf Yes}                   
\\ \hline
\end{tabular}
  }
\end{table*}
\autoref{tab:comparison} qualitatively compares memristive MLNN and in-situ training with earlier works and shows that the proposed work reduces the size of an artificial synapse to one memristor as well as the time to update all the memristors in a crossbar to $\mathbf{O(1)}$.

Our main contributions are as follows: 
\begin{itemize}  
\item[1)] a memristive feed-forward multi-layer neural network architecture, demonstrating resource and space efficiency by utilizing only one memristor for each synapse without  any transistors,
\item[2)] an innovative {\it in-situ} training algorithm based on online gradient descent backpropagation, enabling constant-time weight-matrix update without any additional storage,
\item[3)] a scheme for encoding the input and control signals for the inference and training of the memristive MLNNs,
\item[4)] validation of this memristive MLNN architecture and its {\it in-situ} training method, and the study of its robustness in the presence of (i) device variations, (ii)  stuck-at faults,  (iii) non-linearity in the conductance of memristors (iv) sneak path issues, and (v) scalability of our architecture.
\end{itemize}

 In Section~\ref{background},  the preliminaries of the online gradient descent backpropagation algorithm are outlined. The details of the proposed synapse and neuron in a memristive crossbar and the required peripherals, i.e., encoder and control signals for switches, along with a complete memristive MLNN, are presented in Section~\ref{proposed_idea}. The principles of inference and update operations on a single MCB and then the relevant algorithms for a memristive MLNN are explained in Section~\ref{training}. Simulation results are given in Section~\ref{experiments}, with concluding remarks in Section~\ref{conclusion}. All boldface lowercase and uppercase letters represent vectors and matrices, respectively.

\label{intro}


\section{Background}
\label{background}

Multi-layer neural networks (MLNNs) are trained to learn the data environment with data points. Supervised online gradient descent backpropagation learning algorithm~\cite{Goodfellow-et-al-2016} trains MLNNs where learning is guided by a loss function. An MLNN comprises an input and an output layer of neurons with one or more hidden layers of neurons. The training process involves two stages: forward or feed forward and backpropagation. During forward propagation, the input data is propagated layer by layer from the input layer to the sequence of hidden layers, and finally to the output layer. Next, the gradient or error computed from the loss function is back propagated from the output layer through the hidden layers to the input layer. Each neuron has two states, namely pre-activation and activation. In the pre-activation state, the weighted sum of the inputs is calculated while in the activation state, a non-linear function is applied to this sum. 

\begin{figure}[]
    \centering
\includegraphics[width=\linewidth]{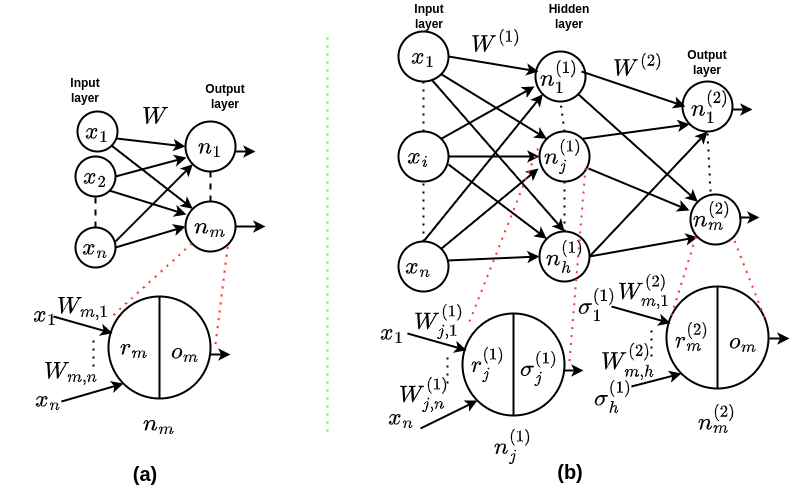}
    \caption{\small{(a) Single-layer neural network (SLNN); (b) Multi-layer neural network (MLNN) with one hidden layer.    
    }}
    \label{fig:SNN}
\end{figure}

For clarity, let us consider a single-layer neural network (SLNN) (Fig.~\ref{fig:SNN}(a)) with $n$ input neurons and $m$ output neurons coupled by an $m \times n$ synaptic weight matrix $W$; there are no hidden layers. An input feature vector $\mathbf{x} \in \mathbb{R}^n$ is fed to the SLNN  to produce the desired output $\mathbf{d} \in \mathbb{R}^m$, where $\mathbf{d}$ is a one-hot vector (for classification problems but may be different for other types ). For a given input $\mathbf{x}$ and its true class label $l$, all the elements of $\mathbf{d}$ are zero except at position $l$. At the pre-activation state, matrix-vector multiplication is performed to obtain the pre-activation vector $\mathbf{r} \in \mathbb{R}^m$ where $\mathbf{r}=\mathbf{Wx}$, having $ r_{j}=\sum _{i=1}^{n} w_{j,i}x_{i}$ for an output neuron $n_j$~\cite{Goodfellow-et-al-2016}.

In order to obtain the forward inferred vector $\mathbf{o}\in \mathbb{R}^m$ at the output layer, a function $f:\mathbb{R}^m \rightarrow \mathbb{R}^m$ is applied to $r_{j}$ of the output neuron $n_{j}$ whose inferred value is the output $o_{j}$. For classification, $f$ is typically {\it soft-max},   so ~$ \forall j=1~to~m$ $o_{j}= f(r_{j}) = \frac{e^{r_{j}}}{\sum_{i=1}^m e^{r_i}}$.  The loss/cost function $L$ defined as $L \triangleq -\sum_{j=1}^md_j.log(o_j)=-log(o_l)$ is {\it cross entropy} where $l$ is the true class label. Other cost functions, for example, mean squared error (MSE) may also be used.

The goal of training is to minimize  $L$ by updating $\mathbf{W}$ which may be chosen randomly at the beginning. The gradient of $L$ w.r.t $\mathbf{r}$, given by the vector $\nabla^L_\mathbf{r}=-\mathbf{(d-o)}$, is calculated and back propagated to update $W$~\cite{Goodfellow-et-al-2016}. By defining $\mathbf{y^{(output)}}\triangleq \mathbf{d}-\mathbf{o}$ as an error vector at the output layer, $\mathbf{\nabla^L_{r}}$ is simplified to $\mathbf{\nabla^L_{r}=-y^{(output)}}$. The $j^{th}$ element $y_{j}$ of $\mathbf{y^{(output)}}$  denotes the error at neuron $n_j$. The gradient of $L$ with respect to the weight matrix $\mathbf{W}$, denoted as $\mathbf{ \nabla^L_{W}}$,  is written in terms of the error vector  $\mathbf{y^{(output)}}$ and input $\mathbf{x}$ as the matrix $\mathbf{ \nabla^L_{W}=-(d-o)x^{T} =-y^{(output)}x^{T}}$~\cite{Goodfellow-et-al-2016}. The weight matrix $\mathbf{W}$ with learning rate $\eta$ is updated as

\begin{equation*}
\mathbf{W(new)}=\mathbf{W(old)}+\mathbf{\Delta W} {\textrm {, where}}
\end{equation*}
\begin{equation}
\label{snn_eq1}
     ~\mathbf{\Delta W \propto - \nabla^L_{W} = -\eta \nabla^L_{W}} =\eta \mathbf{y^{(output)}x^{T}}
\end{equation}
Hence, the change of weight matrix $\mathbf{\Delta W}$ is proportional to the outer product of the vectors $\mathbf{y^{(output)}}$ and $\mathbf{x}$. For a single neuron $n_{j}$, $\Delta w_{j,i}=-\eta \nabla^L_{w_{j,i}}=\eta y_{j}^{(output)}x_{i}$~\footnote{All detailed derivations are in Appendix B (Supplementary Material).}.

Now, consider an MLNN as in Fig.~\ref{fig:SNN}(b) with one hidden layer that contains $h$ hidden neurons. An  $h\times n$ $\mathbf{W^{(1)}}$ connecting the input-hidden layers and an $m\times h$ matrix $\mathbf{W^{(2)}}$ connecting the hidden-output layers are the two weight matrices.
In order to get the activation vector $\bf{\sigma^{(1)}}$ in the activation state at the first/hidden layer, a nonlinear activation function $\mathbf{\sigma}$ is applied on the pre-activation vector $\bf{r^{(1)}}$ of the $1^{st}$ layer to obtain $\bf{\sigma^{(1)}}$, which is is then fed to the output neurons through $\mathbf{W^{(2)}}$. The superscripts in the notation denote layer name/number. The output layer (in this case, the $2^{nd}$ layer) considers the $soft$-$max$ function $f$ as an output function for classification. The four operations $\mathbf{r^{(1)}=W^{(1)}x}$, $\bf{\sigma^{(1)}}={\sigma}(\mathbf{r^{(1)}})$, $\mathbf{r^{(2)}=W^{(2)}\sigma^{(1)}}$, and $\mathbf{o={f(r^{(2)})}}$ are performed in a sequence to infer the output vector $\mathbf{o} \in \mathbb{R}^m $ of the MLNN for a given input $\mathbf{x} \in \mathbb{R}^n$.

The gradient of $L$ with respect to $\mathbf{r^{(2)}}$, i.e., the vector $\mathbf{\nabla^L_{r^{(2)}}}=\mathbf{-y^{(output)}}$, is similar to that for SLNN where $\mathbf{y^{(output)}}$ is the error vector at the output layer. The change in weights of the matrix $\mathbf{W^{(2)}}$ is given by the matrix $\mathbf{\Delta W^{(2)}=-\eta \nabla^L_{W^{(2)}}=\eta y^{(output)}{\sigma^{(1)}}^{T}}$. It is to be noted that the error vectors at the hidden and output layers are different. The gradient of $L$ w.r.t $\mathbf{\sigma^{(1)}}$ is  the vector $\mathbf{\nabla^L_{\sigma^{(1)}}=-(W^{(2)})^Ty^{(output)}=-\delta^{(2)}}$~\cite{Goodfellow-et-al-2016}; and that w.r.t $\mathbf{r^{(1)}}$ is the vector $\mathbf{\nabla^L_{r^{(1)}}=-\delta^{(2)} \odot \sigma'^{(1)}}$~\cite{Goodfellow-et-al-2016} where $\sigma'(\mathbf{x})_{i}=d\sigma(x_{i})/dx_{i}$ and  $\odot $ denotes the element-wise product of two equal-sized vectors. The error vector $\mathbf{y^{(1)}}$ at hidden layer is defined as $\mathbf{y^{(1)}=\delta^{(2)} \odot \sigma'^{(1)}}$, and the gradient of $L$ w.r.t $\mathbf{r^{(1)}}$ in terms of $\mathbf{y^{(1)}}$ is given by $\mathbf{\nabla^L_{r^{(1)}}=-y^{(1)}}$. The gradient of loss function $L$ w.r.t $\mathbf{ W^{(1)}}$ is the matrix  $\mathbf{ \nabla^L_{W^{(1)}}=-y^{(1)}x^{T}}$. Hence, the final update equations for matrices $\mathbf{W^{(2)}}$ and $\mathbf{W^{(1)}}$ are given by
\begin{equation}
\label{mnn_eq11}
    \mathbf{W^{(1)}(new)}=\mathbf{W^{(1)}(old)}+\mathbf{\eta y^{(1)}x^{T}}
\end{equation}
\vspace{-0.75cm}
\begin{equation}
\label{mnn_eq12}
    \mathbf{W^{(2)}(new)}=\mathbf{W^{(2)}(old)}+\mathbf{\eta y^{(output)}{\sigma^{(1)}}^{T}}
\end{equation}


\section{Memristive Neural Network Architecture}
\label{proposed_idea}
\begin{figure*}[]
    \centering
    \includegraphics[width=\linewidth]{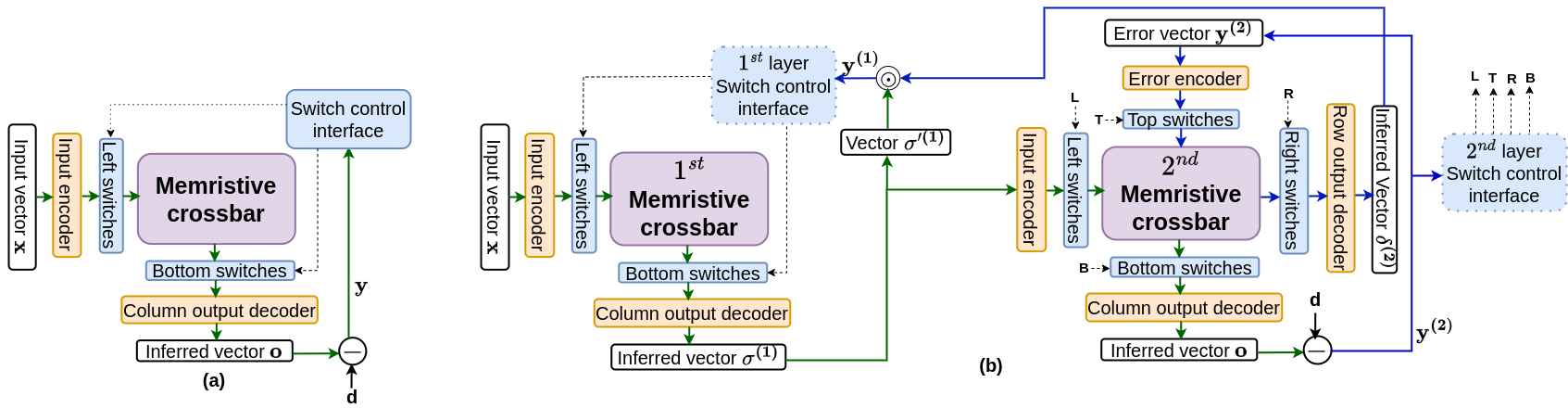}
    \caption{{\small (a) Overview of the proposed memristive SLNN. (b) memristive MLNN with one hidden layer. The operator $\odot$ denotes element-wise multiplication.}}
    \label{fig:overview}
\end{figure*}

An overview of memristive SLNN architecture is shown in Fig.~\ref{fig:overview}(a). Fig.~\ref{fig:overview}(b) illustrates the memristive MLNN with one hidden layer which is obtained by cascading memristive SLNNs. The entire computation mainly consists of forward inference, backward inference, and weight update. The computation of vectors either $\bf{\sigma^{(k)}}$ at $k^{th}$ hidden layer or $\mathbf{o}$ at the output layer is called forward inference. During backward inference at layer $k$, the vector $\delta^{(k)}$ =$(W^{(k)})^Ty^{(k)}$ as defined in Section \ref{background} is calculated while the weight matrix, i.e., the conductance of the memristors in the MCB are updated.

In forward inference, during the period $0$ to $T_{rd}$, the input feature vector $\mathbf{x}$ is encoded into appropriate voltage waveform by the input encoder and fed to the MCB via left switches. This encoded input vector is multiplied with the matrix of the conductance of the memristors in the MCB (in-memory computation) and the result is passed onto the column output decoder via bottom switches. The column output decoder produces either $\mathbf{o}$ or $\bf{\sigma}$ based on layer types.

The vector $\delta^{(k)}$ (Fig. \ref{fig:overview}(b)) is computed during the time period $T_{rd}$ to $2T_{rd}$ only in MLNNs. The error vector $\mathbf{y^{(k)}}$ is encoded into proper voltages by the error encoder and fed back to the $k^{th}$ MCB via top switches. Then in the MCB, the multiplication of the transposed weight matrix and $\mathbf{y^{(k)}}$ is performed and the result is passed, via right switches, onto the row output decoder to produce vector $\mathbf{\delta}^{(k)}$. The error vector $\mathbf{y^{(k-1)}}$ for $(k-1)^{th}$ layer is obtained by element-wise multiplication (denoted by $\odot)$ of $\delta^{(k)}$ and $\sigma'^{(k-1)}$.
While training during the period $2T_{rd}$ to $2T_{rd}+T_{wr}$, the conductances of the memristors in the MCB are updated. 

The arrows in Fig.~\ref{fig:overview} indicate the flow of information. The information flow and update operation are controlled by voltage-controlled switches around the MCB. At $k^{th}$ layer, the switch control interface with $\mathbf{y^{(k)}}$ produces controlling voltage signals to control them. The top and right switches are optional in SLNN and the first layer of MLNN. The details of our proposed memristive neural network architecture are explained in the subsequent sections:
\begin{itemize}
    \item synapses and neurons in a memristive crossbar,
    \item input encoder,
    \item switch control signals for an MCB,
    \item memristive multi-layer neural network.
\end{itemize}

\subsection{Synapses and neurons in a memristive crossbar}
\begin{figure}[h]
    \centering
    \includegraphics[width=\linewidth]{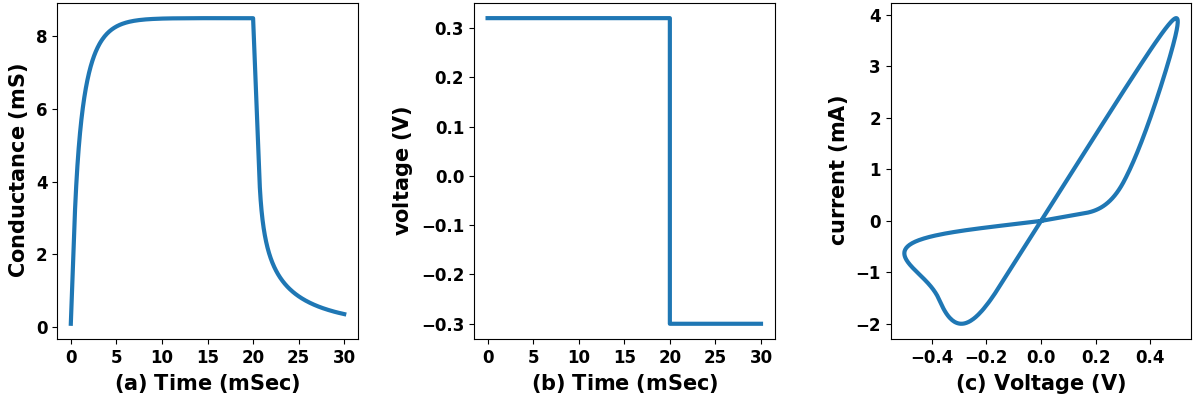}
    \caption{\small{(a) Conductance vs time plot of a memristor; (b) a voltage pulse across a memristor with amplitudes greater than the thresholds; and (c) I-V characteristic curve of a memristor.}}
    \label{fig:mem_conductance}
\end{figure}

A memristor is a two-terminal circuit element having the variable conductance states which depend on a state variable or a set of state variables \cite{Lchua_IEEE_1976}. In this work, only those memristors are considered that show voltage threshold behavior, i.e., changes in their conductances are negligible until they experience voltages greater than the positive threshold voltage ($v^{+}_{th}$), or less than the negative threshold voltage ($v^{-}_{th}$). In this work, the generalized memristor spice model~\cite{TCAD_Yakopcic_2013} is used. In Fig.~\ref{fig:mem_conductance}, the characteristics of a silver chalcogenide-based memristive device~\cite{silver_chalcogenide_memristor_2010_IJCNN} having positive and negative thresholds of +0.16V and -0.15V respectively, are shown.
Fig.~\ref{fig:mem_conductance}(a) shows the variation of conductance of the memristor when voltage pulses, shown in Fig.~\ref{fig:mem_conductance}(b), of amplitude greater than $v^{+}_{th}$ and less than $v^{-}_{th}$ are applied across it. The  I-V curve of the memristor is shown in Fig.~\ref{fig:mem_conductance}(c).

\begin{table}[]
\caption{{\small The specifics of memristive SLNN architecture}}
\label{tab:terms}
\resizebox{\linewidth}{!}{%
\begin{tabular}{|c|c|}
\hline
SLNN Term              & Realisation on an $n\times m$  memristive crossbar \\ \hline
Input neuron $n_{i}$  & Row $i$                                            \\ \hline
Output neuron $n_{j}$ & Column $j$                                         \\ \hline
Synapse $s_{j,i}$ &
  \begin{tabular}[c]{@{}l@{}}Memristor $M_{j,i}$ connecting input neuron $n_{i}$\\  and output neuron $n_{j}$.\end{tabular} \\ \hline
\begin{tabular}[c]{@{}l@{}} Weight $w_{j,i}$ of\\synapse $s_{j,i}$ \end{tabular} &
  \begin{tabular}[c]{@{}l@{}}$w_{j,i}=a\cdot R_{0}(G-G_{j,i})$, includes conductances $G$\\and $G_{j,i}$ of resistor $R$, and memristor $M_{j,i}$ respectively.\end{tabular} \\ \hline
\end{tabular}
}
\end{table}

\begin{figure*}[h]
    \centering
    \includegraphics[width=\linewidth]{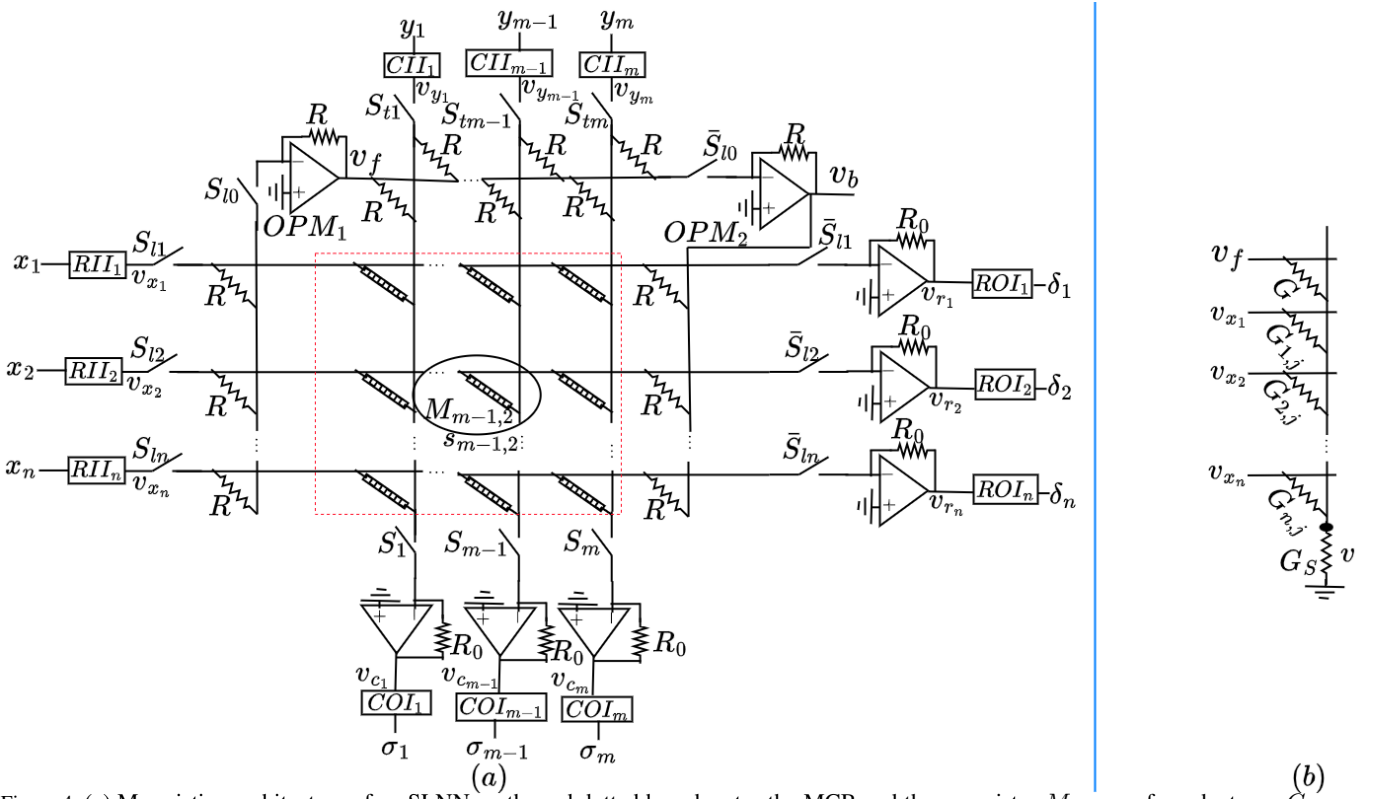}
    \caption{{\small (a) Memristive architecture of an SLNN ---  the red dotted box denotes the MCB and the memristor $M_{m-1,2}$ of conductance $G_{m-1,2}$ is the synapse $s_{m-1,2}$ of weight $w_{m-1,2}$, (b) for column $j$ denoting neuron $n_j$, $G_{j,i}$s, $G$, and $G_{s}$ respectively are the conductances of its $n$ memristors,  the resistor $R$, and the switch $S_j$; $v$ is the voltage across switch $S_{j}$ for $n_{j}$.}}
     \label{fig:cb}
\end{figure*}

The structure, where memristors are arranged in a 2D grid as shown in the red dotted box in Fig.~\ref{fig:cb}(a), is the memristive crossbar (MCB).
In order to implement a memristive SLNN of $n$ input and $m$ output neurons, 
an $n \times m$ MCB,  $(2n+2m)$ resistors, and $(n+m+2)$ op-amps along with a feedback resistor for each, are required. As a well-known practice, a resistor is realized with a pass transistor. The memristive SLNN architecture is shown in Fig.~\ref{fig:cb}(a) and the specifics of its realization on an MCB are in \autoref{tab:terms}.

In  Fig.~\ref{fig:cb}(a), for all $i=1$ to $n$, $x_{i}$ is the $i^{th}$ feature input to the memristive SLNN. The row input interface $RII_{i}$ encodes $x_{i}$ into voltage $v_{x_{i}}$ (see Section~\ref{encoder} below) and feeds it to the $i^{th}$ row via switch $S_{li}$, indicated on the left of the MCB. The inputs to op-amps $OPM_1$ and $OPM_2$ at the top are given via switches $S_{l0}$ and $\bar {S}_{l0}$, and their outputs are $v_{f}$ and $v_{b}$ respectively. The synapse $s_{j,i}$ between input neuron $i$ and output neuron $j$ is represented by memristor $M_{j,i}$.

For all $j=1$ to $m$, $y_{j}$s are the errors to be backpropagated through the crossbar. The column input interface $CII_{j}$ encodes $y_{j}$ into $v_{y_{j}}$ which (only during backpropagation from $T_{rd}$ to $2T_{rd}$, for $T_{rd}$ period) is fed to the $j^{th}$ column via switch $S_{tj}$  at the top of the crossbar. The switch $S_{j}$ is connected at the bottom of the $j^{th}$ column of the MCB. The switch $\bar { S}_{li}$ is placed on the right side of the $i^{th}$ row of the MCB.

The row output interface ($ROI_{i}$) and column output interface ($COI_{j}$) collect the outputs $\delta_{j}$ and $\sigma_{i}$ at the $i^{th}$ row and $j^{th}$ column during backward and forward inferences respectively. During forward inference and update operations, the input voltages are given at the left side of the rows of MCB, whereas the $v_{y_{j}}$ are fed at the top of columns for backward inference. It is to be noted that RIIs, ROIs, CIIs, and COIs form an input encoder, row output decoder, error encoder, and column output decoder, in Fig.~\ref{fig:overview}, respectively.

\subsection{Input encoder}
\label{encoder}
\begin{figure}[]
    \centering
    \includegraphics[width=1\linewidth]{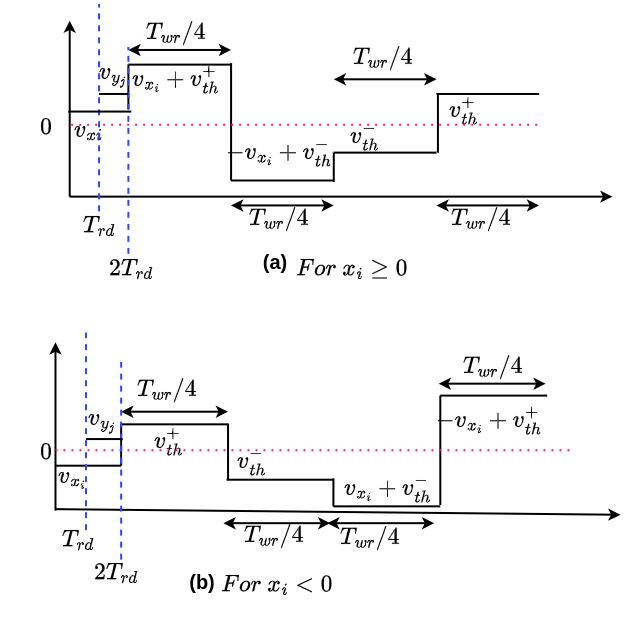}
    \caption{\small { Encoding of input feature $x_{i}$ and error
$y_{j}$ into voltage waveforms for forward and backward propagation, also for updating weights --- (a) $x_{i} \ge 0$, (b) $x_{i} < 0$. For simulation, both rise time and fall times are considered as 1 ns.}}
    \label{fig:input_encoding}
\end{figure}

The input feature and error at neurons are converted into proper voltage waveforms and fed to the memristive crossbar for inferences and updates of the conductances of the memristors in the crossbar during training. 
In Fig.~\ref{fig:cb}(a), the functional peripheral circuits, namely $RII_{i}$ and $CII_{j}$, encode the input $x_{i}$ and error $y_j$ into appropriate voltages respectively, as shown in Fig.~\ref{fig:input_encoding}. For forward inference from $0$ to $T_{rd}$, $RII_{i}$ encodes the input $x_i$ as $v_{x_{i}}=ax_{i}$ where $a$ is a positive constant, and $v_{th}^{-} < v_{x_{i}} < v_{th}^{+}$, as in Fig.~\ref{fig:input_encoding}.

Similarly, $CII_j$ encodes error $y_{j}$ at neuron $n_j$ into $v_{y_{j}},~ 1\leq j\leq m$, and $v_{th}^{-}< v_{y_{j}} < v_{th}^{+}$ to perform backward inference from  $T_{rd}$ to $2T_{rd}$. As $v_{x_{i}}$ and $v_{y_{j}}$ are within the thresholds, there is no change in weight during inferences.

During updates, the conductance of the memristor changes depending on the signs and magnitude of $x_{i}$ and $y_{j}$ as shown in \autoref{tab:update_rules}. The $x_{i}$ part is taken care of by input update voltage. The input voltages for the update are encoded according to $x_{i}$ and exceed the threshold voltages of memristors. For an update operation during the period $2T_{rd}$ to $2T_{rd}+T_{wr}$, the encoding has four possibilities depending on the signs of $x$ and $y$, and the interval $T_{wr}$ has four equal sub-periods with the input update voltage as follows:
\begin{itemize}
    \item $x_{i} \ge 0$ (Fig.~\ref{fig:input_encoding}(a)):
    in the first and the second quarters the input update voltages are $v_{x_{i}}+v_{th}^{+}$ and $-v_{x_{i}}+v_{th}^{-}$, while in the third and the fourth, these are $v_{th}^{-}$ and $v_{th}^{+}$ respectively;
    \item $x_{i} < 0$ (Fig.~\ref{fig:input_encoding}(b)):
    in the first and the second quarters the input update voltages are $v_{th}^{+}$ and $v_{th}^{-}$, while in the third and the fourth, these are $v_{x_{i}}+v_{th}^{-}$ and $-v_{x_{i}}+v_{th}^{+}$ respectively.
\end{itemize}
As $\mathbf{\Delta W}$  depends on input $\mathbf{x}$ by \autoref{snn_eq1}, the voltages during an update operation are also proportional to $\mathbf{x}$.

\subsection{Switch control signals for an MCB}
\label{control}
\begin{figure}[h]
    \centering
\includegraphics[width=1\linewidth]{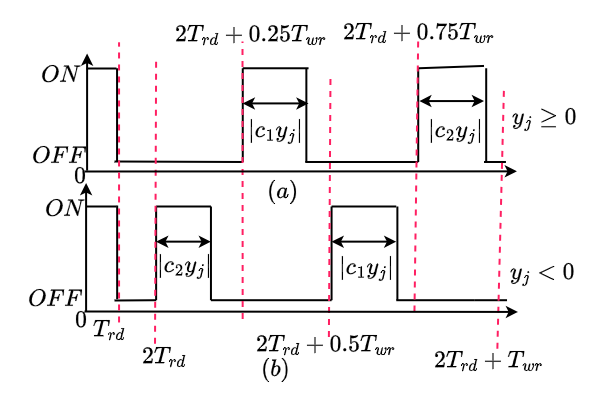}
    \caption{{\small The states of switch $S_{j}$ at the bottom of column $j$ in the MCB of ~Fig.~\ref{fig:cb}(a) --- (a) for $y_{j} \ge 0$; (b) for $y_{j} < 0$. In the simulation, both rise and fall times are considered as 1 ns.}}
    \label{fig:control}
\end{figure}

In Fig. \ref{fig:cb}(a), the signals flow from the left of the rows to the bottom of the columns, and from the top of the columns to the right of the rows of an MCB during forward and backward inferences respectively. Additionally, the change in the conductance of its memristors is controlled by the bottom switches in Fig. \ref{fig:cb}(a). The directions of the signals and conductance update are guided by controlling the ON/OFF states of the voltage-controlled switches in memristive SLNN (Fig~\ref{fig:cb}(a)). The functional block called the switch control interface shown in Fig. \ref{fig:overview}, produces voltages to control the ON/OFF timings of the switches. The switches $S_{li}$, $S_{j}$, $S_{tj}$ and $\bar {S}_{li}$ (for $0 \le i \le n$ and $1 \le j \le m$) are turned ON/OFF to perform the inferences and update operations as follows: 
\begin{itemize}
\item forward inference from $0~to~T_{rd}$ time units: {\it only} the switch $S_{l0}$, switches $S_{li}$ ($1 \le i \le n$) on the left, and switches $S_{j}$ ($1 \le j \le m$) at the bottom are ON; 
\item backward inference (while finding $\delta^{(k)}={W^{(k)}}^Ty^{(k)}$ during backpropagation from $T_{rd}~to~2T_{rd}$ time units): {\it only} switch $\bar{S}_{l0}$, switches $S_{tj}$ ($1 \le j \le m$) at the top, and $\bar {S}_{li}$ ($1 \le i \le n$) on the right are ON; 
\item during update from $2T_{rd}$ to $2T_{rd}+T_{wr}$ time units: switch $S_{l0}$, switch $\bar{S}_{l0}$, switches $S_{tj}$ ($1 \le j \le m$) at the top, and $\bar {S}_{li}$ ($1 \le i \le n$) on the right are OFF. The switches $S_{li}$ ($1 \le i \le n$) on the left are ON but switches $S_{j}$ ($1 \le j \le m$) at the bottom may be ON or OFF depending on error $y_{j}$ at column $j$ during update (Fig.~\ref{fig:control}).
\end{itemize}
For synapse $s_{j,i}$, the $\Delta w_{j,i}=\eta y_{j}x_{i}$ depends on input $x_{i}$ and error $y_{j}$. The $y_{j}$ part is taken care of by the states of switch $S_{j}$. The total update interval $T_{wr}$ is divided into four equal sub-periods. The rules for turning switches $S_{j}$ ON/OFF depending on the sign as well as the magnitude of the error $y_{j}$ at neuron $n_{j}$ are as follows (vide Fig.~\ref{fig:control}) with $c_1$ and $c_2$ being the respective slopes of linearly decreasing and increasing regions of the memristor's conductance (Fig.~\ref{fig:mem_conductance}(a)):
\begin{itemize}
\item  $y_{j}$ $\ge 0$: $S_{j}$ ON for periods proportional to $|c_2y_{j}|$ and  $|c_1y_{j}|$ in the $2^{nd}$ and $4^{th}$ quarters of $T_{wr}$ respectively~(Fig.~\ref{fig:control}(a));
\item $y_{j}<0$: switch $S_{j}$ ON for periods  proportional to $|c_{1}y_{j}|$ and $|c_{2}y_{j}|$ in the $1^{st}$ and $3^{rd}$ quarters of $T_{wr}$ respectively~(Fig.~\ref{fig:control}(b)).
\end{itemize}

\subsection{Memristive multi-layer neural network}
\label{mlnn}
\begin{figure*}[]
     \centering
     \includegraphics[width=\linewidth]{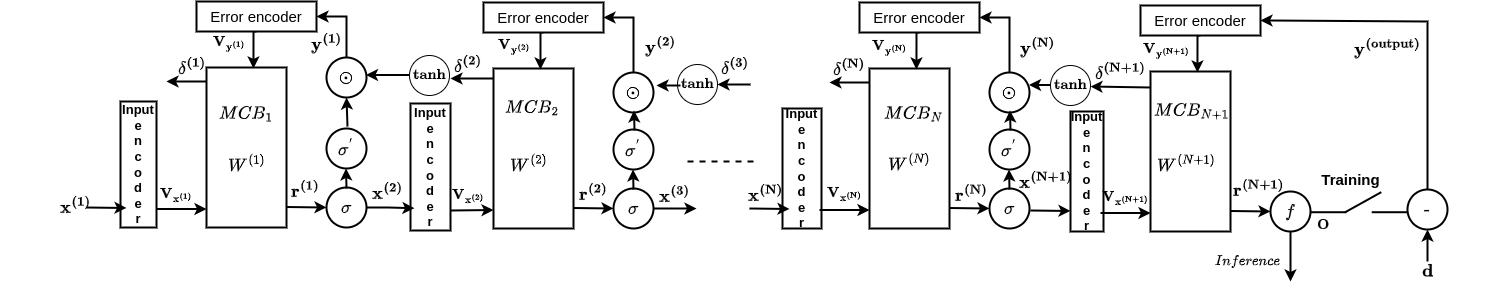}
     \caption{\footnotesize{A schematic of the architecture and training process for the memristive MLNN. The memristive crossbar $MCB_k$ with synaptic weight matrix $W^{(k)}$ connects layers $k$ and $k+1$. The $f$ is the output function at the output layer. All RIIs together form an Input encoder to encode the input vector $\mathbf{x^{(k)}}$ into a proper voltage form using the encoding scheme presented in~Fig.~\ref{fig:input_encoding}. The neural activation function is $\sigma$, and its derivative on the input is $\sigma^{'}$. Element wise product of $\sigma^{'}$ of $(k-1)^{th}$ layer and $\delta^{(k)}$ is denoted by $\odot$, where $\delta^{(k)}={W^{(k)}}^{T}\mathbf{y^{(k)}}$. The $\delta^{(1)}$ is neglected. $y^{(k)}$ is the error vector at layer k.}}
     \label{fig:MLP}
 \end{figure*}

A memristive MLNN of $N$ hidden layers is implemented by cascading $N+1$ memristive crossbars (MCBs) as shown in Fig.~\ref{fig:MLP}, where the output of MCB in layer $l$ is fed to the subsequent MCB for layer $l + 1$. In the memristive MLNN, the input encoder and the error encoder encode input vector $\mathbf{x}$ and error vector $\mathbf{y}$ into $\mathbf{v_{x}}$ and $\mathbf{v_{y}}$ respectively. The input $\mathbf{v_{x}}$ propagates layer by layer from $MCB_{1}$ to $MCB_{N+1}$ while $\mathbf{v_{y}}$ propagates from $MCB_{N+1}$ to $MCB_{1}$. The desired output vector $\mathbf{d}$ is given at the output layer to get the error vector $\mathbf{y^{(output)}}$ only during the training.
The $\delta^{(k)}$ at layer $k$ is rescaled (optional) using $tanh$ function. The arrows in Fig.~\ref{fig:MLP} indicate the flow of data.



\section{In-situ training on a memristive crossbar}
\label{training}
\subsection{Inference and update operations on an MCB}
\label{principle}
In this work, memristors are operated in the linearly increasing and decreasing portions of their $conductance$ vs $time$ plot (Fig.~\ref{fig:mem_conductance}(a)) to maintain all the in-memory activities in the linear domain. Let $G_{max}$ and $G_{min}$ denote the maximum and minimum conductance in the linear region. For neuron $n_{j}$, the conductance of memristor $M_{j,i}$ and switch $S_{j}$ denoted as $G_{j,i}$ and $G_{s}$ respectively typically have the following properties:
\begin{enumerate}
        \item $G_{s}\gg G_{j,i}$  when switch $S_{j}$ is ON;
        \item $G_{s}\ll G_{j,i}$  when switch $S_{j}$ is OFF.
    \end{enumerate}
With these premises, the inference and update on an $MCB$ are illustrated below. 

\subsubsection{Inference}
During forward inference to get the vector $\sigma$, the states of the switches~in Fig.~\ref{fig:cb}(a) in anti-clockwise order are 
\hspace{0 in} $S_{li}: ON$, $S_{j}: ON$, $\bar{S}_{li}: OFF$ and $S_{tj}: OFF$; for $0\le i \le n$ and $1\leq j\leq m$ from $0~to~T_{rd}$ time units. Fig.~\ref{fig:cb}(b) shows the conductance equivalent circuit of the neuron $n_j$, in which the memristor $M_{j,i}$, resistor $R$ of $OPM_{1}$, and switch $S_{j}$ of column $j$ have been replaced by the corresponding conductance $G_{j,i}$, $G$, and $G_{s}$ respectively, where $1\leq i\leq n$. It is noted that resistors and switches can be realized with transistors. We compute the voltages across memristors of the neuron $n_{j}$ during inference and update to observe their effects on conductance $G_{j,i}$ of $M_{j,i}$. Recall that $RII_{i}$ encodes $x_i$ as $v_{x_{i}}=ax_{i}$ (refer Fig.~\ref{fig:input_encoding}) such that $v_{th}^{-} < v_{x_{i}}< v_{th}^{+}$ for inference. It is necessary to know the voltages $v$ across switch $S_{j}$ and $(v_{x_i}-v)$ across memristor $M_{j,i}$ (refer Fig.~\ref{fig:cb}(b)) to perform the inference and update on memristive MCB. With $ G=\frac{1}{R}=\frac{G_{max}+G_{min}}{2}$, we have
\vspace{-0.5cm}
\begin{equation*}
    \begin{split}
        & (v_{f}-v)G+\sum_{i=1}^{n}(v_{x_{i}}-v)G_{j,i}=vG_{s}
    \end{split}
\end{equation*}
\vspace{-0.5cm}
\begin{equation}
    \label{volt_v}
    v=\frac{v_{f}G+ \sum_{i=1}^{n} v_{x_{i}}G_{j,i}}{G_{s}+ G+ \sum_{i=1}^{n}G_{j,i}}
\end{equation}
As $S_{j}$ is ON,  $G_{s} \gg G~$ and $G_{s} \gg G_{j,i}$, for $i=1~to~n$.  Thus, if $G_{s} \gg \sum_{i=1}^{n}G_{j,i}$, from \autoref{volt_v} we get
    \begin{equation*}
             v= \frac{v_{f}G+ \sum_{i=1}^{n} v_{xi}G_{j,i}}{G_{s}}
             =v_{f} \frac{G}{G_{s}}+\sum_{i=1}^{n} v_{x_{i}} \frac{G_{j,i}}{G_{s}}
    \end{equation*}
    \vspace{-0.75cm}
    \begin{equation}
        \label{v_equals}
        v\approx 0
    \end{equation}
    Therefore, $M_{j,i}$ experiences $v_{x_{i}}-v \approx v_{x_{i}}$, the input voltage to $i^{th}$ row which lies in the threshold range of $M_{j,i}$.  This implies no change in $G_{j,i}$ of $M_{j,i}$, and in the weight $w_{j.i}~(=aR_{0}(G-G_{j,i})$) of synapse $s_{j,i}$. Again, referring to Fig.~\ref{fig:cb}(a), the voltage $v_{f}=-\sum_{\forall i} \frac{R}{R}v_{x_{i}}=-\sum_{\forall i}v_{x_{i}}$ and the output $v_{c_{j}}$ of op-amp at bottom of $j^{th}$ column is
\begin{equation*}
    \label{output}
        v_{c_{j}} =-\left[\sum_{i=1}^{n}(R_{0}G_{j,i}\times v_{x_{i}})+R_{0}G\times v_{f}\right]
        \end{equation*}
    \begin{equation*}
        v_{c_{j}} =-\sum_{i=1}^{n}R_{0}(G_{j,i}-G)v_{x_{i}}\\
        =\sum_{i=1}^{n}aR_{0}(G-G_{j,i})\frac{v_{x_{i}}}{a} 
\end{equation*}
Defining the weights $w_{j,i}=aR_{0}(G-G_{j,i})$, we have $v_{c_{j}}=\sum_{i=1}^{n}w_{j,i}x_{i}$ where $x_{i}=\frac{v_{x_{i}}}{a}$. The weight $w_{j,i}$ is positive if $G>G_{j,i}$ else non-positive. It is to be noted that for $j=1~to~m$, all $v_{c_j}$ together form a pre-activated vector $\mathbf{r=Wx}$, i.e., the weight matrix $\mathbf{W}$ with $w_{j,i}$ as its entries multiplied by the input vector $\mathbf{x}$. Therefore, the matrix-vector multiplication is performed within the period $T_{rd}$ and hence $\mathbf{O(1)}$ time. The $COI_j$ produces activated output $\sigma_j$ over $r_j$ of $n_j$.

During backpropagation, to get vector $\bf{\delta}$ consisting of $\delta_{i}$, for ~$i=1~to~n$, needs the states of the switches~(in Fig.~\ref{fig:cb}(a)) to be
\hspace{0 in} $S_{li}:OFF$, $S_{j}:OFF$, $\bar{S}_{li}:ON$ and $S_{tj}:ON$; $0\leq i\leq n$ and $1\leq j\leq m$ from $T_{rd}~to~2T_{rd}$ time units. \\
This is simply an inference operation where the inputs $v_{y_{j}}$  for $j: 1 \leq j \leq m$ are fed at the top of the MCB, and the vector $\bf{\delta}$ are inferred by $ROI$s interfaces.
This operation is also performed within $T_{rd}$ time period, i.e., $\mathbf{O(1)}$. 

\subsubsection{Update}
\label{write}
By \autoref{snn_eq1},  the change in weight of synapse $s_{j,i}$ is given by $\Delta w_{j,i}=\eta y_{j}x_{i}$, which depends on the signs and values of the input $x_{i}$ and the error $y_{j}$ in four possible ways, as shown in \autoref{tab:update_rules}. 
\begin{table}[h]
    \centering
    \caption{{\small The rules to update the weight $w_{j,i} =a\cdot R_{0}(G-G_{j,i})$ --- change in conductance $G_{j,i}$ of memristor $M_{j,i}$ with changes in input $x_{i}$ and error $y_{j}$.} }
    \label{tab:update_rules}
    \resizebox{\linewidth}{!}{%
    \begin{tabular}{|c|c|c|c|c|}
    \hline
       Input $x_{i}$  &  Error  $y_{j}$ &  \begin{tabular}[c]{@{}l@{}}Weight\\change $\Delta w_{j,i}$ \end{tabular} & \begin{tabular}[c]{@{}l@{}}   Conductance\\ change $G_{j,i}$ \end{tabular} & \begin{tabular}[c]{@{}l@{}} Updating period decided by \\ $v_{x_{i}}$ (Fig.~\ref{fig:input_encoding}) and $S_{j}$ (Fig.~\ref{fig:control}) \end{tabular} \\
      \hline 
       $+{ve}$ & $+{ve}$ & $+{ve}$ & $\downarrow $ & $2T_{rd}+0.25T_{wr}$ to $2T_{rd}+0.5T_{wr}$\\
       \hline
        $+{ve}$ & $-{ve}$ & $-{ve}$ & $\uparrow $ & $2T_{rd}$ to $2T_{rd}+0.25T_{wr}$\\
       \hline
        $-{ve}$ & $+{ve}$ & $-{ve}$ & $\uparrow $ & $2T_{rd}+0.75T_{wr}$ to $2T_{rd}+T_{wr}$\\
       \hline
        $-{ve}$ & $-{ve}$ & $+{ve}$ & $\downarrow $ & $2T_{rd}+0.5T_{wr}$ to $2T_{rd}+0.75T_{wr}$\\
       \hline
    \end{tabular}
    }
\end{table}

 The weight $w_{j,i} (=a\cdot R_{0}(G-G_{j,i})$) of synapse $s_{j,i}$ increases or decreases if the value of $G_{j,i}$ of $M_{j,i}$ decreases or increases respectively. 
 During the update from $2T_{rd}$ to $2T_{rd}+T_{wr}$ time units, the switch states are:\\
\hspace{0 in} $S_{l0}$ and $\bar{S}_{l0}:OFF$, $S_{li}:ON$, $\bar{S}_{li}:OFF$ and $S_{tj}:OFF$, $1\leq i\leq n$ and $1\leq j\leq m$ . \\ Therefore, $v_{f}=0$. At $n_{j}$, assuming the input feature $x_{i} \ge 0$, then input update voltage $v^{*}_{x_{i}}$ at $row_{i}$ has to be $v_{th}^{+}<v^{*}_{x_{i}}<2v_{th}^{+}$ and $2v_{th}^{-}<v^{*}_{x_{i}} <v_{th}^{-}$ in the first and second quarters of $T_{wr}$ as per input encoding in Fig.~\ref{fig:input_encoding}. Referring to Fig.~\ref{fig:control} and Section~\ref{control}, if error $y_{j} \ge 0$ at $n_{j}$, then switch $S_j$ is $ON$ up to the period proportional to $|y_j|$ in the second and the fourth quarters of $T_{wr}$. Under this condition, upto a time period of  $|c_1y_{j}|$ in the second quarter of $T_{wr}$, the memristor $M_{j,i}$ experiences a negative voltage $v^{*}_{x_{i}}$ (=$-v_{x_{i}}+v_{th}^{-}$) and $G_{j,i}$ decreases~($\bigtriangleup G_{j,i} \propto -x_{i}y_{j}$) that results in the increase of $w_{j,i}$ ($\propto x_{i}y_{j}$). In the fourth quarter of $T_{wr}$ (refer Fig.~\ref{fig:control} and Fig.~\ref{fig:input_encoding}), $M_{j,i}$ experiences a positive voltage but within the threshold range, so no change in $G_{j,i}$ occurs. 

However, the salient question is that if $S_{j}$ of neuron $n_j$ is OFF during $T_{wr}$, will $G_{j,i}$ change? The answer is NO. The reason is as follows. 

If $S_{j}$ is OFF, then $G_{s} \ll G$ and $G_{s} \ll G_{j,i}$  for $i=1, 2, \ldots ,n$ and $v_{f}=0$, then from \autoref{volt_v}
        \begin{equation*}
        \begin{split}
            & v= \frac{\sum_{i=1}^{n} v^{*}_{x_{i}}G_{j,i}}{G+ \sum_{i=1}^{n}G_{j,i}}
        \end{split}
    \end{equation*}

Suppose we get $v = v^{*}$. The input update voltage $v^{*}_{x_{i}}$ at $i^{th}$ row is either positive or negative but not both in any quarter of $T_{wr}$ as shown in~Fig.~\ref{fig:input_encoding}. So the $v^{*}$ will be either $2v^{-}_{th} < v^{*} < v^{-}_{th}$, or $v^{+}_{th} < v^{*} < 2v^{+}_{th}$ in any quarter of $T_{wr}$. Therefore, each $M_{j,i}$ of column $j$ experiences a voltage across it in  the range $v_{th}^{-}< v^{*}_{x_{i}}-v = v^{*}_{x_{i}}-v^{*} <v_{th}^{+}$. Therefore, no change in $G_{j,i}$ results in no update in $w_{j,i}$. Similarly, for the other three combinations of signs of the input $x_{i}$ and error $y_{j}$ as given in \autoref{tab:update_rules}, the $G_{j,i}$ of $M_{j,i}$ is updated accordingly resulting $w_{j,i}$ is modified correctly. Therefore, there is no conductance alteration in column $j$ if $S_{j}$ is OFF.

In the linear region of the conductance $G_{j,i}$ of $M_{j,i}$ (Fig.~\ref{fig:mem_conductance}(a)),  $\Delta G_{j,i}$ is proportional to the product of the update voltages and the time duration of this voltage. During the update, the change $\Delta G_{j,i}$ is jointly taken care of by input voltage $v^{*}_{x_{i}}$ and the time duration ($\propto |y_j|$) for which the bottom switches $S_{j}$ are ON. Since $\Delta G_{j,i}$ is proportional to the product of input $x_i$ and time $|y_j|$ ($\propto |y_j|.x_{i}$), the weight $w_{j,i}$ is updated as per \autoref{snn_eq1}.

Since each $M_{j,i}$ is updating its $G_{j,i}$ independently for an input $x_{i}$ and error $y_{j}$, therefore all $G_{j,i}$s of a crossbar are updated in $\mathbf{O(1)}$ time. Therefore, the weight updates that involve the outer product of the vectors and addition operations (\autoref{snn_eq1}), are carried out in constant time with an MCB. With these inferences and update operations on an MCB, the {\it in-situ} training of the memristive MLNN is presented next.

\subsection{Training of a memristive MLNN}
\label{in_situ_algo}
The proposed {\it in-situ} supervised algorithms for the training of a memristive MLNN are based on gradient descent backpropagation. Fig.~\ref{fig:MLP} is a flow diagram of the training. The data set is divided into training and test sets. 
\subsubsection{Inference on memristive MLNN}
This operation is performed during both training and testing. The inference vector $\mathbf{o}$ at the output layer from the input $\mathbf{x}$ is obtained with Algorithm~\ref{algo:forwardPass}. The accuracy is calculated on the test set.
\begin{algorithm}[h]
    { \small
    \caption{Inference operation on a memristive MLNN with $N$ hidden layers}
    \label{algo:forwardPass}
    \KwData{Input vector $\mathbf{x}$ from training data set;}
    \KwResult{Anticipated output vector $\mathbf{o}$;}
     $\mathbf{x^{(1)}} = \mathbf{x}$\;
    \For{$k=1$ to $N$ + $1$}{
         Apply Input ENCODER comprising row input interfaces (RIIs), to encode $\mathbf{x^{(k)}}$ into equivalent voltages $\mathbf{v_{x^{(k)}}}$\;
        Apply $\mathbf{v_{x^{(k)}}}$ to $MCB_{k}$\;
        
        for $p=0~to~n$, keep all switches $S_{lp}$ (left) ON and $\bar{S}_{lp}$ (right) OFF in $MCB_{k}$ for inference duration of $T_{rd}$\;
        
       for $q=1~to~m$, keep  all switches $S_{q}$ (bottom) $ON$  and $S_{tq}$ (top) $OFF$ in $MCB_{k}$ for inference duration of $T_{rd}$ \;
       
        Assign voltage $\mathbf{v_{c}^{(k)}}$ of $k^{th}$ $MCB_{k}$ to $\mathbf{r^{(k)}}$\;
        \If{$k<$ $N$ + $1$}{
            $\mathbf{x^{(k+1)}} = \mathbf{\sigma(r^{(k)})}$, the input for $MCB_{k+1}$ by applying the activation function to $\mathbf{r^{(k)}}$\;
            }
        \Else{
            $\mathbf{o} = \mathbf{f(r^{(k)})}$, the predicted values at the output layer.           
            }
        }     
    }
\end{algorithm}
\subsubsection{Update on memristive MLNN} During training, the weight matrix in the MCB is updated. The error $\mathbf{y^{(output)}}$ at the output layer is the difference between the forward inferred vector $\mathbf{o}$ and the labeled output vector $\mathbf{d}$. The $\mathbf{y^{(output)}}$ is back propagated and the synaptic weights are adjusted layer by layer as described in Algorithm~\ref{algo:weightUpdate}. During backpropagation, the crossbar $MCB_{i}$ is used to get the vectors $\mathbf{\delta^{(i)}}$ at the $i^{th}$ layer in $\mathbf{O(1)}$ time as described in Algorithm~\ref{algo:weightUpdate} (line  5), which accelerates the backpropagation. 
\begin{algorithm}[h]
    {\small
    \caption{Update operation for the weight matrices of a memristive MLNN with $N$ hidden layers}
    \label{algo:weightUpdate}
    \KwData{Input vector $\mathbf{x}$, anticipated output vector $\mathbf{o}$ of Algorithm 1,  labeled output vector $\mathbf{d}$ for $\mathbf{x}$;

    }
    \KwResult{Proper weight adjustment of each synapse in the memristive MLNN;}
$\mathbf{y^{(N+1)}}$ = $\mathbf{y^{(output)}}$, the error at the output layer with $\mathbf{o}$ obtained from Algorithm.~\ref{algo:forwardPass} and $\mathbf{d}$\; 
    \For{$k = N + 1$ to $1$}{
Get error voltages $\mathbf{v_{y^{(k)}}}$ from error ENCODER by feeding $\mathbf{y^{(k)}}$ to it\;
        Apply $\mathbf{v_{y^{(k)}}}$ to $MCB_{k}$\;
        \For{ $p=0~to~n$, $q=0~to~m$} {keep  switches $\Bar{S}_{lp}$ and $S_{tq}$  ON, and $S_{lp}$ and $S_q$ OFF for the time period $T_{rd}$  to get $\mathbf{\delta^{(k)}}=\mathbf{{(W^{(k)})}^T}\mathbf{y^{(k)}}$\;
        }
         Update the crossbar $MCB_{k}$ using update rules in \autoref{tab:update_rules} (Section \ref{write})\;
        \If{$k>1$}{
            $\mathbf{y^{(k-1)}} =  tanh(\mathbf{\delta^{(k)}}) \mathbf{ \odot} \mathbf{\sigma'^{(k-1)}}$.
            }
      }
}
\end{algorithm}

From Algorithms~\ref{algo:forwardPass} and \ref{algo:weightUpdate}, it is to be noted that for a layer, the three operations of matrix-vector multiplication, $\delta^{(2)}$ operation, and MCB update are performed within periods of $T_{rd},~T_{rd}$, and $T_{wr}$ respectively, all in $\mathbf{O(1)}$ time.


\section{Experimental Results}
\label{experiments}
The proposed memristive SLNNs and MLNN were simulated using the open source LTSpice XVII simulator, running on an Ubuntu 20.04 LTS environment with an 8-core 1.6GHz Intel Core {\it i5} processor and 8GB RAM. The datasets for classification, network type with the number of features in each layer, number of crossbars and their sizes, and resistor $R_{0}$ are given in~\autoref{tab:circuit_parameters}. The values of $T_{rd}$ and $T_{wr}$ are chosen as $10 \mu s $ and $1 ms$ respectively. For all simulations carried out with memristors, memristive crossbars, resistors, Opamps, switches, and functional blocks such as $RII_{i},~ROI_{j}$, etc., the wire resistance has been assumed to be negligible. Further, we simulated a memristive MLNN with MNIST dataset~\cite{lecun-mnisthandwrittendigit-2010} in Python to demonstrate the scalability of the architecture.
\begin{table}[]
\caption{Neural network parameters used in our LTSpice simulation}
\label{tab:circuit_parameters}
\resizebox{\linewidth}{!}{%
\begin{tabular}{|c|c|c|c|c|}
\hline
\multirow{2}{*}{Data set} &
  \multirow{2}{*}{Network type, \# features} &
  \multirow{2}{*}{\#MCBs with size} &
  Memristor model~\cite{silver_chalcogenide_memristor_2010_IJCNN} &
  Memristor model~\cite{Miller_EDL_2010} \\ \cline{4-5} 
              &                                                &                                     & $R_{0}$ in $\Omega$ & $R_{0}$ in $\Omega$  \\ \hline
NASA Asteroid~\cite{nasa:2018} & Single layer, $20 \rightarrow 1$               & One; $21 \times 1$                  & $100$   & $2k$    \\ \hline
\begin{tabular}[c]{@{}c@{}}Breast Cancer\\ Wisconsin\end{tabular}~\cite{bcwiris:2019} &
  Single layer, $30 \rightarrow 1$ &
  One; $31 \times 1$ &
  $100$ &
  $15k$ \\ \hline
XOR           & Multi-layer , $2 \rightarrow 2 \rightarrow 2$  & Two; $3  \times 2$ and $3 \times 2$ & $1k$    & ------  \\ \hline
IRIS~\cite{bcwiris:2019}         & Multi-layer , $4 \rightarrow 4  \rightarrow 3$ & Two; $5  \times 4$ and $5 \times 3$ & $1k$    & $15k$   \\ \hline
\end{tabular}
}
\end{table}

For a dataset, if $n$ and $m$ are the number of features and classes, then $n+1$ (along with a bias) and $m$ are the input and output neurons at the input and output layers respectively. The number of hidden layers and hidden neurons are decided empirically by experiments. A sigmoid function was used at the output layer in SLNNs. The sigmoid and $tanh$ were used as activation functions at the hidden layers in memristive MLNNs for IRIS and XOR datasets respectively. The output function chosen in MLNNs was $soft$-$max$ while cross entropy was the cost function. The input data $\mathbf{x}$ belongs to that neuron/class which has the highest $soft$-$max$ value. 

The Spice models of silver chalcogenide~\cite{silver_chalcogenide_memristor_2010_IJCNN} and anodic titania~\cite{Miller_EDL_2010} based memristors have been employed. These memristor spice models are in~\cite{TCAD_Yakopcic_2013}. The values of the parameters\footnote{Glossary in \autoref{tab:parameters_descriptions} of Appendix \ref{parameters_description}} in~\autoref{tab:parameter_silver} used in general Spice model~\cite{TCAD_Yakopcic_2013} are to match the characterizations of silver chalcogenide~\cite{silver_chalcogenide_memristor_2010_IJCNN} and anodic titania~\cite{Miller_EDL_2010} based memristors. From the simulation, the average energy consumed at a synapse to perform both read and write operations is $1.589 \mu J$.
\begin{figure}[h]
    \centering
    \includegraphics[width=0.9\linewidth]{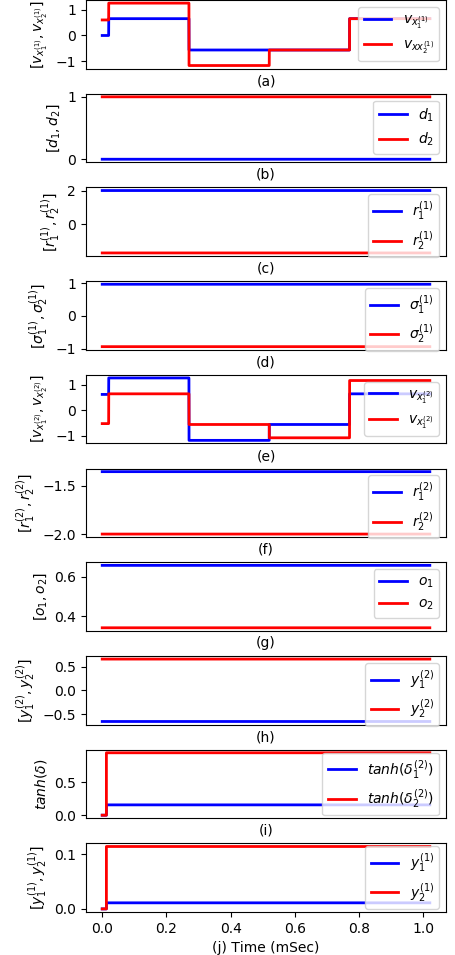}
    \caption{{\small For an input data point in $[1, 0]$ of the XOR dataset, the plots (a) to (j) represent the training steps of our memristive MLNN. In (d), the output $[\sigma^{(1)}_{1}, \sigma^{(1)}_{2}]$ of the hidden layer is the input $[\mathbf{x}^{(2)}_{1}, \mathbf{x}^{(2)}_{2}]$ to the output layer. In (i), $tanh(\delta)=[tanh(\delta^{(2)}_{1}),tanh(\delta^{(2)}_{2}]$.}}
    \label{fig:xor}
\end{figure}

{\it An illustration of the steps:} A memristive MLNN on the XOR dataset with network structure mentioned in \autoref{tab:circuit_parameters} was simulated to explain the training steps (Fig.~\ref{fig:MLP}) and the results are presented in Fig.~\ref{fig:xor}. The four inputs are $[0,0]$, $[0,1]$, $[1,0]$, and $[1,1]$ whereas the respective output vectors are $[1,0]$, $[0,1]$, $[0,1]$, and $[1,0]$. As an example, for input $[0,1]$ with constant $a=0.6$, the encoded input voltages $[v^{(1)}_{x_{1}},v^{(1)}_{x_{2}}]$ during forward inference and update are shown Fig.~\ref{fig:xor}(a) and the corresponding desired output vector $[d1, d2]$ is in Fig.~\ref{fig:xor}(b). The weights (memristors' conductance) are initialized randomly. The weighted sums $[r^{(1)}_{1}, r^{(1)}_{2}]$ (equivalent to $v_{c_1}$ and $v_{c_2}$ in MCB in Fig.~\ref{fig:cb}) during forward inference of the hidden layer is in Fig.~\ref{fig:xor}(c). The output $[\sigma^{(1)}_{1}, \sigma^{(1)}_{2}]$ in the hidden layer is in Fig.~\ref{fig:xor}(d). The $[\sigma^{(1)}_{1}, \sigma^{(1)}_{2}]$ ($ \equiv [\mathbf{x}^{(2)}_{1}, \mathbf{x}^{(2)}_{2}]$) is encoded for the next layer, as presented in Fig.~\ref{fig:xor}(e). Fig.~\ref{fig:xor}(f) shows  the weighted sums $[r^{(2)}_{1}, r^{(2)}_{2}]$ of output layer. The inferred output vector [$o_{1}$, $o_{1}$] is in Fig.~\ref{fig:xor}(g). The error $[y^{(2)}_{1}, y^{(2)}_{2}]$ at output layer is shown in Fig.~\ref{fig:xor}(h). The vector $[\delta^{(2)}_{1},\delta^{(2)}_{2}]$ is calculated with $MCB_{2}$ during backward inference and rescaled with tanh function as $[tanh(\delta^{(2)}_{1}), tanh(\delta^{(2)}_{2})]$ and shown in Fig.~\ref{fig:xor}(i). Fig.~\ref{fig:xor}(j) shows the error at the first layer.

\begin{table}[h]
\centering
\caption{Parameters used in  Spice models for memristors in ~\cite{TCAD_Yakopcic_2013} 
}
\label{tab:parameter_silver}
\resizebox{\linewidth}{!}{%
\begin{tabular}{|c|c|c|c|c|c|c|c|c|c|c|c|c|}
\hline
Memristor Parameter $\rightarrow$& \textbf{$a_1$} & \textbf{$a_2$} & \textbf{$b$} & \textbf{$a_p$} & \textbf{$a_n$} & \textbf{$X_p$} & \textbf{$X_n$} & \textbf{$V_p$} & \textbf{$V_n$} & \textbf{$\alpha_{p}$} & \textbf{$\alpha_{n}$} & \textbf{$\eta$} \\ \hline
 Memristor model~\cite{silver_chalcogenide_memristor_2010_IJCNN}      & 0.17           & 0.17           & 0.05         & 4000             & 4000             & 0.3            & 0.5            & 0.16           & 0.15           & 1          & 5       & 1               \\ \hline
 Memristor model~\cite{Miller_EDL_2010}      & 1.4           & 1.4           & 0.05         & 16            & 11             & 0.3            & 0.5            & 0.65           & 0.56           & 1.1          & 6.2       & -1               \\ \hline
\end{tabular}
}
\end{table}

\subsubsection*{Simulation with memristor model in ~\cite{silver_chalcogenide_memristor_2010_IJCNN} }In the spice model~\cite{silver_chalcogenide_memristor_2010_IJCNN} for silver chalcogenide memristors, the threshold values $v_{th}^{+}$, and $v_{th}^{-}$ are $0.16V$ and $-0.15V$ respectively. The minimum and maximum conductance are $0.255 mS$ and $8.5 mS$ respectively while $G_{min}$ and $G_{max}$ in the linear region of the memristor's conductance are $3.18 mS$ and $6.38 mS$ respectively. The conductance $G_{j,i}$ of memristor $M_{j,i}$ was initialized in the range $\{4.4 mS, 5 mS\}$. The memristive SLNNs and MLNNs were trained with algorithms \ref{algo:forwardPass} and \ref{algo:weightUpdate} on NASA Asteroid, Breast Cancer Wisconsin, and IRIS data sets with $90.43\%,~98.59\%,~and~98.22\%$ accuracies respectively.
\begin{figure}[]
    \centering
    \includegraphics[width=\linewidth]{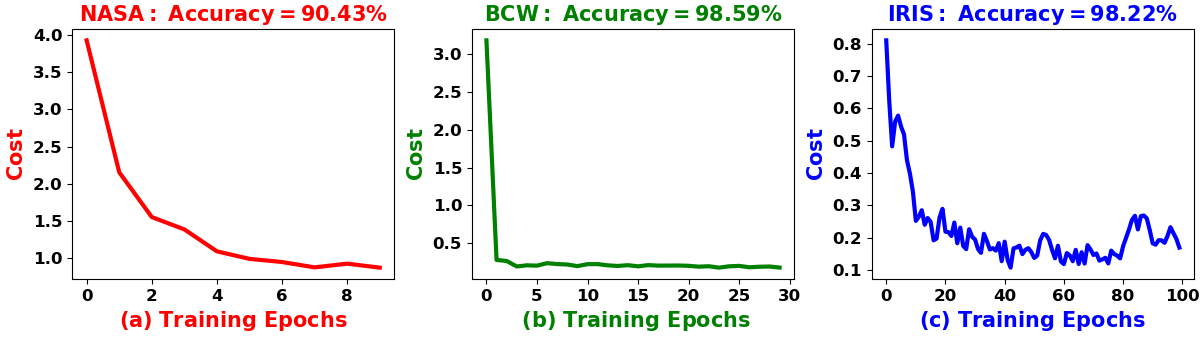}
    \caption{{\small The cost function during training of memristive neural networks constructed with silver chalcogenide based memristors~\cite{silver_chalcogenide_memristor_2010_IJCNN} for (a) NASA Asteroid (b) Breast Cancer Wisconsin (c) IRIS datasets.}}
    \label{fig:cost_silver}
\end{figure}
The corresponding cost functions during training are shown in Figures. \ref{fig:cost_silver}(a), (b), and (c) respectively for these three data sets. For each dataset, the number of training epochs as indicated has been chosen empirically to avoid overfitting.

\subsubsection*{Simulation with memristor model in~\cite{Miller_EDL_2010}} Our proposed memristive ANN architecture was also tested with anodic titania-based spice model~\cite{Miller_EDL_2010} of a memristor.  Its threshold voltages $v_{th}^{+}$ and $v_{th}^{-}$ are $0.65V$ and $-0.56V$ respectively. The minimum and maximum conductances are $1 mS$ and $70 mS$, respectively. The linear section of its conductance plot has  $28 mS$ and $48 mS$ as the values of $G_{min}$ and $G_{max}$, respectively. The initialization range for the conductance of a memristor was between $35 mS$ and $41 mS$. The memristive SLNNs and MLNNs were trained using proposed in-situ algorithms on NASA Asteroid, Breast Cancer Wisconsin, and IRIS data sets with $90.40\%,~97.54\%,~and~98.22\%$ accuracies respectively. 
\begin{figure}[]
    \centering
    \includegraphics[width=\linewidth]{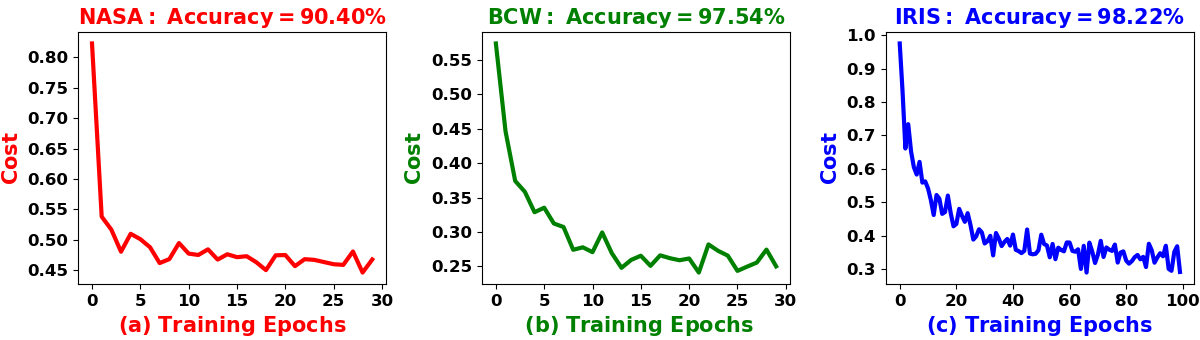}
    \caption{{\small The cost function during training of memristive neural networks constructed with anodic titania-based memristors~\cite{Miller_EDL_2010} for (a) NASA Asteroid (b) Breast Cancer Wisconsin (c) IRIS datasets.}}
    \label{fig:cost_anodic}
\end{figure}
Figures \ref{fig:cost_anodic}(a), (b), and  (c) are the costs during training. 

\subsubsection*{Performance of proposed architecture} The performance is compared with similar works in~\cite{santlal2022, shahar15,shahar16, yan2020NN} and the classification accuracy with both memristor models~\cite{silver_chalcogenide_memristor_2010_IJCNN} and \cite{Miller_EDL_2010} are presented in \autoref{tab:accuracy}, where the one in~\cite{silver_chalcogenide_memristor_2010_IJCNN}  gives a slightly better value. 
\begin{table}[h]
\caption{Comparison of classification accuracy (\%) with prior similar works~\cite{santlal2022, shahar15, shahar16,yan2020NN}.}
\label{tab:accuracy}
\resizebox{\linewidth}{!}{%
\begin{tabular}{|c|c|c|c|c|c|cc|}
\hline
\multirow{2}{*}{Data set} &
  \multirow{2}{*}{Network structure} &
  \multirow{2}{*}{\cite{santlal2022}} &
  \multirow{2}{*}{\cite{shahar15}} &
  \multirow{2}{*}{\cite{shahar16}} &
  \multirow{2}{*}{\cite{yan2020NN}} &
  \multicolumn{2}{l|}{Proposed work with memristor} \\ \cline{7-8} 
 &
   &
   &
   &
   &
   &
  \multicolumn{1}{c|}{Model~\cite{silver_chalcogenide_memristor_2010_IJCNN}} &
  Model~\cite{Miller_EDL_2010} \\ \hline
NASA Asteroid &
  Single-layer &
  89.04 &
  $\mathbf{\dots}$ &
  $\mathbf{\dots}$ &
  $\mathbf{\dots}$ &
  \multicolumn{1}{c|}{90.43} &
  90.40 \\ \hline
\begin{tabular}[c]{@{}c@{}}Breast Cancer\\ Wisconsin\end{tabular} &
  Single-layer &
  90.14 &
  98.7 &
  97 &
  97.72 &
  \multicolumn{1}{c|}{98.59} &
  97.54 \\ \hline
IRIS &
  \begin{tabular}[c]{@{}c@{}}Multi-layer with\\ one hidden layer\end{tabular} &
  99.11 &
  97.33 &
  84.33 &
  88.43 &
  \multicolumn{1}{c|}{98.22} &
  98.22 \\ \hline
\end{tabular}
}
\end{table}
\begin{table*}[]
\caption{{\small Memristor parameters  increased or decreased from original values in \autoref{tab:parameter_silver} in order to obtain $10\%$ variation of I-V characteristic in memristor models of ~\cite{silver_chalcogenide_memristor_2010_IJCNN, Miller_EDL_2010}; these values are taken from ~\cite{TCAD_Yakopcic_2013}.}}
\label{tab:variation_parameter}
\resizebox{\linewidth}{!}{%
\begin{tabular}{|lll|l|l|l|l|l|l|l|l|l|l|l|}
\hline
\multicolumn{3}{|l|}{{\color[HTML]{000000} Memristor Parameter $\rightarrow$}} &
  {\color[HTML]{000000} $a_1$} &
  {\color[HTML]{000000} $a_2$} &
  {\color[HTML]{000000} $b$} &
  {\color[HTML]{000000} $a_p$} &
  {\color[HTML]{000000} $a_n$} &
  {\color[HTML]{000000} $X_p$} &
  {\color[HTML]{000000} $X_n$} &
  {\color[HTML]{000000} $V_p$} &
  {\color[HTML]{000000} $V_n$} &
  {\color[HTML]{000000} $\alpha_{p}$} &
  {\color[HTML]{000000} $\alpha_{n}$} \\ \hline
\multicolumn{2}{|l|}{{\color[HTML]{000000} }} &
  {\color[HTML]{000000} Model~\cite{silver_chalcogenide_memristor_2010_IJCNN}} &
  {\color[HTML]{000000} 0.153} &
  {\color[HTML]{000000} 0.153} &
  {\color[HTML]{000000} 0.045} &
  {\color[HTML]{000000} 2680} &
  {\color[HTML]{000000} 2680} &
  {\color[HTML]{000000} 0.18462} &
  {\color[HTML]{000000} 0.3077} &
  {\color[HTML]{000000} 0.104848} &
  {\color[HTML]{000000} 0.098295} &
  {\color[HTML]{000000} 0.145} &
  {\color[HTML]{000000} 0.725} \\ \cline{3-14} 
\multicolumn{2}{|l|}{\multirow{-2}{*}{{\color[HTML]{000000} Decreased values}}} &
  {\color[HTML]{000000} Model~\cite{Miller_EDL_2010}} &
  {\color[HTML]{000000} 1.26} &
  {\color[HTML]{000000} 1.26} &
  {\color[HTML]{000000} 0.045} &
  {\color[HTML]{000000} 9.888} &
  {\color[HTML]{000000} 6.798} &
  {\color[HTML]{000000} 0.2139} &
  {\color[HTML]{000000} 0.3565} &
  {\color[HTML]{000000} 0.594165} &
  {\color[HTML]{000000} 0.511896} &
  {\color[HTML]{000000} 0} &
  {\color[HTML]{000000} 0} \\ \hline
\multicolumn{2}{|l|}{{\color[HTML]{000000} }} &
  {\color[HTML]{000000} Model~\cite{silver_chalcogenide_memristor_2010_IJCNN}} &
  {\color[HTML]{000000} 0.187} &
  {\color[HTML]{000000} 0.187} &
  {\color[HTML]{000000} 0.055} &
  {\color[HTML]{000000} 5924} &
  {\color[HTML]{000000} 5924} &
  {\color[HTML]{000000} 0.42363} &
  {\color[HTML]{000000} 0.70605} &
  {\color[HTML]{000000} 0.217696} &
  {\color[HTML]{000000} 0.20409} &
  {\color[HTML]{000000} 2.115} &
  {\color[HTML]{000000} 10.575} \\ \cline{3-14} 
\multicolumn{2}{|l|}{\multirow{-2}{*}{{\color[HTML]{000000} Increased values}}} &
  {\color[HTML]{000000} Model~\cite{Miller_EDL_2010}} &
  {\color[HTML]{000000} 1.54} &
  {\color[HTML]{000000} 1.54} &
  {\color[HTML]{000000} 0.055} &
  {\color[HTML]{000000} 32.16} &
  {\color[HTML]{000000} 22.11} &
  {\color[HTML]{000000} 0.57903} &
  {\color[HTML]{000000} 0.96505} &
  {\color[HTML]{000000} 0.6994} &
  {\color[HTML]{000000} 0.60256} &
  {\color[HTML]{000000} 1.9855} &
  {\color[HTML]{000000} 11.191} \\ \hline
\end{tabular}
}
\end{table*}

\subsection{Variability analysis for the proposed architecture}
In order to analyze the memristor's device variation issues, we added variations in memristors' spice models \cite{silver_chalcogenide_memristor_2010_IJCNN,Miller_EDL_2010} and evaluated the performance of memristive MLNN on IRIS dataset. In order to add $\nu \%$ variations in the I-V characteristics of the memristors, the fitting parameters were adjusted until the change in I-V characteristic exceeded $\nu \%$~\cite{TCAD_Yakopcic_2013}. The $10\%$ change in I-V characteristic is measured using the total average difference between the altered I-V characteristic with parameters in \autoref{tab:variation_parameter} and the initial I-V characteristic with parameters in \autoref{tab:parameter_silver}~\cite{TCAD_Yakopcic_2013}.

\begin{table*}[]
\caption{Accuracy and F1-Score with $10\%$ I-V curve variation added to randomly chosen memristors in the MLNN on IRIS dataset.}
\label{tab:variation_result}
\resizebox{\linewidth}{!}{%
\begin{tabular}{|c|cccc|cccc|}
\hline
\multirow{3}{*}{\begin{tabular}[c]{@{}l@{}} $\%$ of memristors\\ having variation \end{tabular}} & \multicolumn{4}{l|}{\begin{tabular}[c]{@{}l@{}}Variation added by decreasing memristor\\ parameters\end{tabular}} & \multicolumn{4}{l|}{\begin{tabular}[c]{@{}l@{}}Variation added by increasing memristor\\ parameters\end{tabular}} \\ \cline{2-9} 
                                                                                   & \multicolumn{2}{c|}{Model~\cite{silver_chalcogenide_memristor_2010_IJCNN}}                                         & \multicolumn{2}{c|}{Model~\cite{Miller_EDL_2010}}                    & \multicolumn{2}{c|}{Model~\cite{silver_chalcogenide_memristor_2010_IJCNN}}                                         & \multicolumn{2}{c|}{Model~\cite{Miller_EDL_2010}}                   \\ \cline{2-9} 
                                                                                   & \multicolumn{1}{l|}{Accuracy}   & \multicolumn{1}{l|}{F1-Score}    & \multicolumn{1}{l|}{Accuracy}   & F1-Score    & \multicolumn{1}{l|}{Accuracy}   & \multicolumn{1}{l|}{F1-Score}    & \multicolumn{1}{l|}{Accuracy}   & F1-Score   \\ \hline
10                                                                                 & \multicolumn{1}{l|}{$97.33\%$}  & \multicolumn{1}{l|}{$96.214\%$}  & \multicolumn{1}{l|}{$96.44\%$}  & $94.667\%$  & \multicolumn{1}{l|}{$98.22\%$}  & \multicolumn{1}{l|}{$97.598\%$}  & \multicolumn{1}{l|}{$97.33\%$}  & $96.02\%$  \\ \hline
20                                                                                 & \multicolumn{1}{l|}{$97.33\%$}  & \multicolumn{1}{l|}{$96\%$}      & \multicolumn{1}{l|}{$96.44\%$}  & $94.667\%$  & \multicolumn{1}{l|}{$98.22\%$}  & \multicolumn{1}{l|}{$97.598\%$}  & \multicolumn{1}{l|}{$97.33\%$}  & $96.209\%$ \\ \hline
30                                                                                 & \multicolumn{1}{l|}{$97.33\%$}  & \multicolumn{1}{l|}{$96.214\%$}  & \multicolumn{1}{l|}{$96.44\%$}  & $94.667\%$  & \multicolumn{1}{l|}{$98.22\%$}  & \multicolumn{1}{l|}{$97.598\%$}  & \multicolumn{1}{l|}{$96.44\%$}  & $95.03\%$  \\ \hline
\end{tabular}}
\end{table*}


In order to analyze the effects of variation on the performance of our memristive MLNN, we have chosen randomly $10\%$, $20\%$, and $30\%$ of the memristors and added $10\%$ I-V curve variation in these selected memristors.
Simulation results in \autoref{tab:variation_result} demonstrate that the variations in memristors have not affected the performance significantly.

\subsection{Robustness of the architecture against stuck-at-a-conductance state} The robustness against faulty memristors is crucial for maintaining performance. To address this, several simulation experiments were carried out when a certain percentage (tested for 1\%, 5\%, 10\%, and 20\% of memristors randomly chosen) of the memristors in the crossbars were stuck-at-a-conducting state and unable to update themselves during training. These experiments were done with the spice model~\cite{silver_chalcogenide_memristor_2010_IJCNN} to assess the effect in the above circumstance. 
\begin{table*}[h]
\centering
    \caption{{\small Classification accuracy of our architecture after training with faulty memristors and comparison with  ~\cite{santlal2022}.}}
    \label{tab:robust1}
\begin{tabular}{|c|ccccc|ccccc|}
\hline
  & \multicolumn{5}{c|}{\begin{tabular}[c]{@{}c@{}}Memristive neural network\\ architecture in~\cite{santlal2022} \end{tabular}}    & \multicolumn{5}{c|}{\begin{tabular}[c]{@{}c@{}}Proposed memristive  neural \\ network architecture\end{tabular}} \\ \hline
\begin{tabular}[c]{@{}c@{}}\% of memristors randomly chosen \\ as stuck-at a conductance state\end{tabular}                                       & \multicolumn{1}{c|}{0 \%}  & \multicolumn{1}{c|}{1 \%}  & \multicolumn{1}{c|}{5 \%}  & \multicolumn{1}{c|}{10 \%} & 20 \% & \multicolumn{1}{c|}{0 \%}   & \multicolumn{1}{c|}{1 \%}   & \multicolumn{1}{c|}{5 \%}   & \multicolumn{1}{c|}{10 \%} & 20 \% \\ \hline
\begin{tabular}[c]{@{}c@{}}Classification accuracy (\%) for \\ NASA Asteroid data set\end{tabular}          & \multicolumn{1}{c|}{89.04} & \multicolumn{1}{c|}{85.68} & \multicolumn{1}{c|}{79.97} & \multicolumn{1}{c|}{82.89} & 72.83 & \multicolumn{1}{c|}{90.43}  & \multicolumn{1}{c|}{{90.43}}  & \multicolumn{1}{c|}{{91.86}}  & \multicolumn{1}{c|}{{92.24}} & {89.39} \\ \hline
\begin{tabular}[c]{@{}l@{}}Classification accuracy (\%) for\\ Breast Cancer Wisconsin data set\end{tabular} & \multicolumn{1}{l|}{90.14} & \multicolumn{1}{l|}{83.8}  & \multicolumn{1}{l|}{82.04} & \multicolumn{1}{l|}{81.34} & 79.23 & \multicolumn{1}{l|}{{98.59}}  & \multicolumn{1}{l|}{{98.59}}  & \multicolumn{1}{l|}{{98.59}}  & \multicolumn{1}{l|}{{99.65}} & {98.24} \\ \hline
\begin{tabular}[c]{@{}l@{}}Classification accuracy (\%) for \\ IRIS data set\end{tabular}                   & \multicolumn{1}{l|}{99.11} & \multicolumn{1}{l|}{99.11} & \multicolumn{1}{l|}{96.44} & \multicolumn{1}{l|}{94.67} & 80.44 & \multicolumn{1}{l|}{{98.22}}  & \multicolumn{1}{l|}{{98.22}}  & \multicolumn{1}{l|}{{97.33}}  & \multicolumn{1}{l|}{{97.33}} & { 98.22} \\ \hline
\end{tabular}
\end{table*}
After training, the accuracy has not been affected much as evident from \autoref{tab:robust1} and compared with a similar study in~\cite{santlal2022}. This demonstrates that compared to~\cite{santlal2022}, the architecture and {\it in-situ} training are more robust to the presence of faulty memristors. The authors \cite{li_2018_nature_communication} performed fault analysis on a memristive MLNN with one hidden layer (2M type synapse and $\mathbf{O(n)}$ MCB update time) on the MNIST dataset ($8\times8$ image size, i.e., $64$ input neurons) where the accuracy was $91.7\%$ with $11\%$ memristors being stuck. 

\subsection{Sneak path immunity}
The sneak-path problem is a common challenge in 1M memristive crossbar arrays. Various methods have been proposed to either mitigate sneak-path effects during functional operation~\cite{Shevgoor2015ICCD} or to exploit the sneak-path current for testing purposes~\cite{sneakTCAD}. The proposed model exhibits inherent robustness against this issue, as the sneak-path voltages across all unintended memristors remain below $v_{\text{th}}$, resulting in no undesired updates under all conditions.

\begin{figure}
    \centering
    \includegraphics[width=\linewidth]{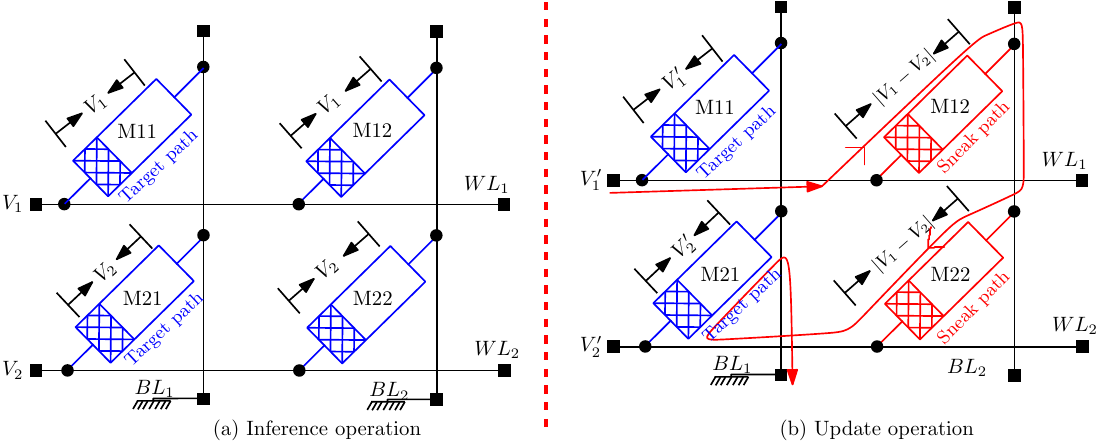}
    \caption{(a) During inference operations, voltages $v_1$ \text{and} $v_2$ are applied across $M_{11}, M_{12}$ and $M_{21}, M_{22}$, respectively, where $v_1 \geq v_2 \quad \text{and} \quad v_1, v_2 \leq v_{\text{th}}$. (b) During update operation, voltages $v'_1$ and $v'_2$ are applied across $M_{11}$ and $M_{21}$, respectively, where $v'_1=v_1+v_{ht}$ and $v'_2=v_2+v_{ht}$. An undesired sneak path voltage $|v'_1-v'_2|=|v_1-v_2|\leq v_{th}$ is appeared across $M_{12}$ and $M_{21}$, respectively, resulting no effect.}
    \label{fig:sneak}
\end{figure}

To demonstrate this behavior, a $2 \times 2$ memristive crossbar array is considered, as illustrated in Fig.~\ref{fig:sneak}. During the inference phase Fig.~\ref{fig:sneak}(a), each memristor is subjected to either $v_1$ or $v_2$, both of which are below the threshold voltage $v_{\text{th}}$. As a result, no change in memristor conductance occurs, effectively preventing sneak path-induced disturbances.
During the update phase Fig.~\ref{fig:sneak}(b), assume that only neuron $n_1$ (corresponding to the first column) undergoes synaptic modification, while neuron $n_2$ does not. During update operation, voltages $v'_1$ and $v'_2$ are applied across $M_{11}$ and $M_{21}$, respectively, where $v'_1=v_1+v_{\text{th}}$ and $v'_2=v_2+v_{\text{th}}$. An undesired sneak path voltage $|v'_1-v'_2|=|v_1-v_2|\leq v_{\text{th}}$ is appeared across $M_{12}$ and $M_{21}$, respectively, which remains below the threshold. This ensures that no unintended updates occur, thereby effectively mitigating the sneak path issue.

\subsection{Effects of non-linearity on the performance}
The memristor's conductance changes non-linearly. In order to examine the effect of the non-linearity in conductance (refer Fig.~\ref{fig:mem_conductance}(a)) on performance, four experiments were performed where memristors were initialized (i) only in the linear region, (ii) in both linear and non-linear region, (iii) only in the lower non-linear region, and (iv) only in the upper non-linear region. 
\begin{figure}[]
    \centering
    \includegraphics[width=\linewidth]{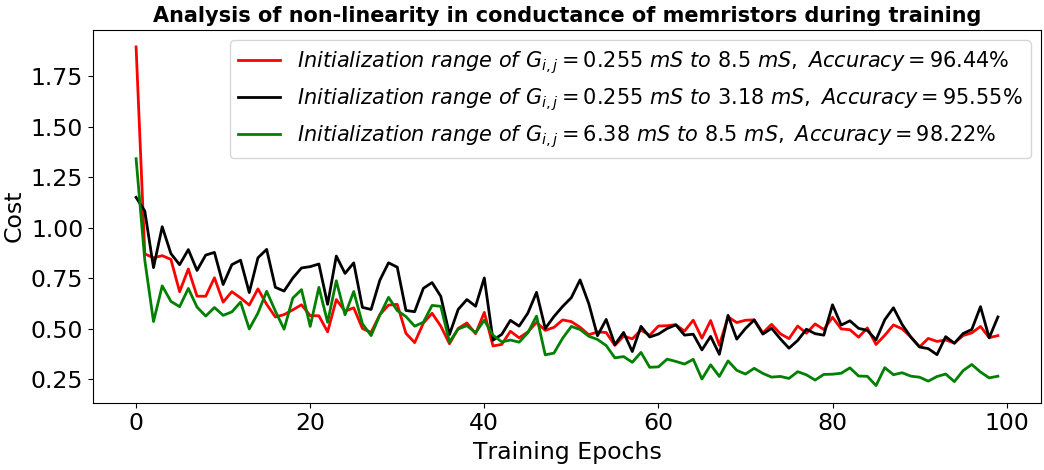}
    \caption{{\small Effects of nonlinearity in the conductance of a memristor (Fig.~\ref{fig:mem_conductance}(a)) on the performance for IRIS dataset; the conductance of the memristors \cite{silver_chalcogenide_memristor_2010_IJCNN} of the crossbar are initialized in the range (ii) \{$0.225 mS, 8.5 mS$\} (red), (iii) only lower non-linear \{$0.225 mS, 3.18 mS$\} (black), and (iv) only upper non-linear \{$6.38 mS, 8.5 mS$\} (green). For range (i) only linear conductance region, the accuracy shown in Fig. \ref{fig:cost_silver}(c) is 98.22\%. }}
    \label{fig:cost_nonlinear_silver}
\end{figure}
The analysis was performed with memristor model~\cite{silver_chalcogenide_memristor_2010_IJCNN} that was trained and tested with the  IRIS data set. For case (i), we get $98.22\%$ accuracy as shown in Fig.~\ref{fig:cost_silver}(c). The red plot in Fig.~\ref{fig:cost_nonlinear_silver} shows the loss during training with an accuracy of $96.44\%$ when the memristors are initialized in the range of \{$0.225mS,~ 8.5mS$\} (linear and non-linear regions of conductance in Fig.~\ref{fig:mem_conductance}(a)). The black plot in Fig.~\ref{fig:cost_nonlinear_silver} shows the loss during training with accuracy of $95.55\%$ when memristors were initialized in the lower non-linear range \{$0.225mS,~ 3.18mS$\}. When the memristors were randomly initialized only in the upper non-linear range \{ $6.38 mS$, $8.5 mS$\}, it gave $98.22\%$ accuracy and the cost is shown in the green plot in Fig.~\ref{fig:cost_nonlinear_silver}. In summary, the results of these experiments establish that the training in the non-linear region on this architecture with its {\it in-situ} training algorithm does not affect the accuracy drastically compared to that with the linear region only.

\subsection{Scalability of the proposed architecture} We performed digit classification of MNIST~\cite{lecun-mnisthandwrittendigit-2010} data set to test the scalability of the proposed architecture. The training and testing data sets had 60000 and 10000 images of shape $28 \times 28$ pixels respectively and each pixel is considered as a feature. The LTspice simulator becomes very slow for large MCBs. Hence the spice netlists of memristive MLNN for the MNIST dataset are not suitable.
\begin{figure}[h]
    \centering
    \includegraphics[width=0.8\linewidth]{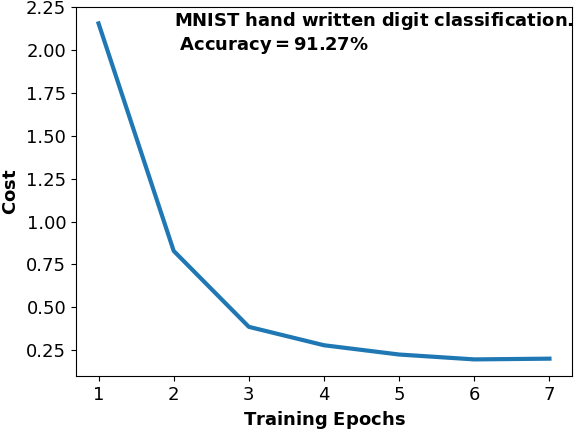}
    \caption{{\small The cost function (cross-entropy) during training of proposed architecture on MNIST data set for digit classification.}}
    \label{fig:cost_mnist}
\end{figure}
\begin{table}[h]
\centering
\caption{\small{Comparative Performance on MNIST Dataset}}
\label{mnist_accuracy}
\resizebox{\linewidth}{!}{%
\begin{tabular}{|c|c|c|c|}
\hline
\textbf{Reference}         & \textbf{Accuracy (\%)} & \textbf{Dataset}        & \textbf{Key Details}                         \\ \hline
\cite{he2023Neurocomputing} & 97.16                 & MNIST                   & \begin{tabular}[c]{@{}l@{}} 4000 epochs, Matlab simulation \\ mathematical model of network    \end{tabular}         \\ \hline
\cite{yan2020NN}           & 82.2                  & MNIST                   & 2T1M synapses, SimElectronics, MATLAB  \\ \hline
\cite{yilmaz2024TETCI}      & 92.39                 &\begin{tabular}[c]{@{}l@{}} Binarised MNIST \\ ($28 \times 28$) \end{tabular} & 1M synapses, O(n²) update time, Neurosim            \\ \hline
\cite{Yilmaz2023NCA}       & 94.0                  & \begin{tabular}[c]{@{}l@{}} Binarised MNIST \\ ($20 \times 20$) \end{tabular} & 1M synapses, O(n²) update time, Neurosim            \\ \hline
\textbf{Our Work}                   & 91.27                 & MNIST                   &1M synapse, $\mathbf{O(1)}$ update time, Python          \\ \hline
\end{tabular}
}
\end{table}
There is also a limit of 1024 on the number of nodes in a subcircuit that this simulator can handle. Because of the above limitations, the proposed algorithms for memristive MLNN with two hidden layers ($784 \rightarrow 397 \rightarrow 204 \rightarrow 10$ and total $394887$ synapses) were simulated in Python. The cost during training is shown in Fig.~\ref{fig:cost_mnist}.
The proposed model achieves $91.27\%$ accuracy on MNIST using Python with just 2 hidden layers and 7 epochs. \autoref{mnist_accuracy} summarizes previous works with varying accuracy with different synapse configurations, update times, and simulation environments.

\
\subsubsection*{Discussion}
Compared to~\cite{santlal2022}, this architecture uses 50\% fewer memristors, employing a 1M MCB without transistors or synapse-controlling devices. It addresses resilience to faults from stuck-at-conductance states, not covered in~\cite{santlal2022}, and, to the best of our knowledge, is the first to explore the impact of non-linearity in memristor conductance. We study performance variations due to memristor imperfections in detail, unlike~\cite{santlal2022}. While~\cite{santlal2022} only considered conductance increases, our model supports both increases and decreases based on input and error, utilizing separate encoding and distinct control signals for positive and negative inputs and errors.


This work focuses on memristive MLNN and its related aspects. Compared with MLNNs, the data flow in convolutional neural networks (CNNs) is different. Moreover, major operations like stride, pooling during forward propagation, and full convolution operation during backpropagation are performed in CNNs. The future direction will be to modify the proposed memristive architecture for in-situ trainable CNNs that will be efficient in area, energy, and training latency.

\section{Conclusion}
\label{conclusion}
A novel architecture of memristive multi-layer neural networks with an efficient {\it in-situ} training algorithm is proposed here. The training algorithm based on gradient descent back propagation updates all the memristors of a crossbar in $\mathcal{O}(1)$ time. Here, one memristor suffices for a single synapse. Furthermore, it is demonstrated that the accuracy of the MLNN is not significantly impacted, even if some of the memristors (tested for 1\%, 5\%, 10\%, and 20\% of the memristors) are stuck at a conducting state.
Experimental analysis revealed that the variation and non-linearity in the conductance of memristors do not have a notable impact on the accuracy of the MLNN. At a synapse, the average energy required to execute read and write operations is $1.589 \mu J$. The comprehensive energy analysis is being investigated and will be provided separately.



\bibliographystyle{IEEEtran}
 \bibliography{main.bib}

\appendix

\section{Memristor parameters and relevant derivations for Neural Networks}
\subsection{Glossary of parameters for memristor model}
\label{parameters_description}
The parameters of the  memristor model~\cite{TCAD_Yakopcic_2013} for LTSpice used in Section \ref{experiments} are described in \autoref{tab:parameters_descriptions}.

\begin{table}[h]
\begin{center}
\caption{ {\small Parameters in Spice Model for Memristor ~\cite{TCAD_Yakopcic_2013}}}
\label{tab:parameters_descriptions}
 \resizebox{\linewidth}{!}{%
\begin{tabular}{|l|l|}
\hline
\textbf{Parameter}                      & \textbf{Relation to Physical Behaviors}                                                     \\ \hline
$a_1$ and $a_2$                & \begin{tabular}[c]{@{}l@{}} Closely related to the thickness of the\\ dielectric layer in a memristor device, as more electrons\\ can tunnel through a thinner barrier leading to an increase\\ in conductivity.\end{tabular}                                                                                                                                     \\ \hline
b                              & \begin{tabular}[c]{@{}l@{}} Determines how much curvature is seen in\\ the I-V curve relative to the applied voltage. This relates\\ to how much of the conduction in the device is Ohmic and\\ how much is due to the tunnel barrier.\end{tabular}                                                                                                                \\ \hline
$A_p$ and $A_n$                & \begin{tabular}[c]{@{}l@{}}These control the speed of ion (or filament)\\ motion. This could be related to the dielectric material used\\ since oxygen vacancies have different mobility depending\\ which metal-oxide they are contained in.\end{tabular}                                                                                                          \\ \hline
$V_p$ and $V_n$                & \begin{tabular}[c]{@{}l@{}}These represent the threshold voltages. There \\ may be related to the number of oxygen vacancies in\\ a device in its initial state. A device with more oxygen\\ vacancies should have a larger current draw that may lead\\ to a lower switching threshold if switching is assumed to\\ be based on the total charge applied\end{tabular} \\ \hline
$\alpha_p$, $\alpha_n$, $x_p$ and $x_n$ & \begin{tabular}[c]{@{}l@{}}These determine where the state variable motion is\\ no longer linear, and  determine the degree to which\\ the state variable motion is dampened. This could be related to\\ the electrode metal used on either side of the dielectric film\\ since the metals chosen may react to the oxygen vacancies\\ differently.\end{tabular}           \\ \hline
\end{tabular}
}
\end{center}
\end{table}

\end{document}